\documentclass[structabstract]{aa}  
\usepackage{longtable}
\usepackage{graphicx}
\usepackage{latexsym}
\usepackage{amssymb}
\usepackage{lscape}
\usepackage{txfonts}
\begin{document}
   \title{The complete census of optically selected AGNs in the Coma Supercluster: the dependence of AGN activity on the local environment}

   \author{Giuseppe Gavazzi \inst{1}, Giulia Savorgnan  \inst{1}, Mattia Fumagalli \inst{2}
          }
   \institute{
   	    Dipartimento di Fisica G. Occhialini, Universit\`a di Milano- Bicocca, Piazza della Scienza 3, 20126 Milano, Italy\\
             \email{giuseppe.gavazzi@mib.infn.it} 
   	     \and
	     Leiden Observatory, Leiden University, P.O. Box 9513, 2300 RA Leiden, The Netherlands.\\
             \email{fumagalli@strw.leidenuniv.nl}
	     }
              \date{Accepted 18/07/2011}

 
  \abstract
   {}
   {To investigate the dependence of the occurrence of active galactic nuclei (hereafter AGNs)
   on local galaxy density, 
   we study the nuclear properties of $\sim$ 5000 galaxies in the Coma Supercluster whose 
   density spans two orders of magnitude from the sparse filaments to the cores of rich clusters.
   }
   {
   We obtained optical spectra of the nuclei of 283 galaxies using the 1.5m Cassini telescope of Bologna observatory.  
   Among these galaxies, 177 belong to the Coma Supercluster and are added to the 4785 spectra available from SDSS (DR7) to
   fill-in the incomplete coverage by SDSS of luminous galaxies. 
   We perform a spectral classification of the nuclei of galaxies in this region (with a completeness of 98\% at $r\leq$17.77),
   classifying the nuclear spectra in six classes: three of them (SEY, sAGN, LIN) refer to active galactic nuclei and the remaining three 
   (HII, RET, PAS) refer to different stages of starburst activity. We perform this classification 
   as recommended by Cid Fernandes and collaborators using
   the ratio of $\lambda$ 6584 [NII] to $\rm H\alpha$ lines and the equivalent width (EW) of $\rm H\alpha$ (WHAN diagnostic), 
   after correcting the last quantity by $1.3~\AA$ for underlying absorption. 
   }
   {
   We find that 482 (10\%) of 5027 galaxies host an active galactic nucleus (AGN): 
   their frequency strongly increases with increasing luminosity of the parent galaxies,
   such that 32\% of galaxies with $\rm Log (L_i/L_{\odot})\geq 10.2~ $ harbor an active galactic nucleus at their interior.
   In addition to their presence in luminous galaxies, AGNs are also found in red galaxies 
   with $<g-i> \simeq 1.15 \pm 0.15$ mag.
   The majority of SEY and sAGN (strong AGNs) are associated with luminous late-type (or S0a) galaxies, 
   while LIN (weak AGNs)
   and RET ("retired": nuclei that have experienced a starburst phase in the past 
   and are now ionized by their hot evolved low-mass stars), 
   are mostly found among E/S0as.
   The number density of AGNs, HII region-like, and retired galaxies is found to anti-correlate with the local density of galaxies, 
   such that their frequency drops by a factor of two near the cluster cores, while the frequency of galaxies containing passive nuclei increases by the same amount 
   towards the center of rich clusters. The dependence of AGN number density on the local galaxy density 
   is greater than the one implied by morphology segregation alone.
   }
   {}

%
\authorrunning{Gavazzi et al.}
\titlerunning{Nuclear spectroscopy of galaxies} 
\maketitle

\section{Introduction}

It appears that
1.5m class telescopes, especially those ones located in zones of marginal climatic conditions, 
have lost much of their $raison~d'etre$, particularly after the advent of the Sloan Digital Sky Survey (SDSS).
They however maintain a crucial didactic value for the training of young astronomers
and provided that they are employed for suitable projects, can still make some significant niche contribution.
One appropriate project in this respect is, in our opinion, the determination of the spectral properties of 
nuclei of bright galaxies,
as part of a general study to perform a complete census of active galactic nuclei (AGNs).
A spectral resolution on the order of R=1000 in the red channel (the blue-arm is more penalized by
the relative blindness of CCDs) is sufficient to measure in a matter of a few minutes, even in non-photometric conditions, 
some crucial line ratios in the nuclear regions of galaxies, owing to their high surface brightness.\\
In this spirit, one of us (G.G.) invited yearly his first year master students to participate in some observing runs
at the 1.5m Loiano telescope of the Observatory of Bologna (It). 
Unsurprisingly, many nights were affected by poor weather, although an amazing wealth of
nuclear spectra (283) of nearby galaxies was accumulated, suitable for detecting any AGNs harbored in their interior.\\
A large amount of data (177) were gathered for the region of the Coma Supercluster, on which this work is focused.
Following a fortunate run of 7/8 photometric nights in spring 2011
and the appearance of Mahajan et al. (2010), who investigated the active galactic nucleus (AGN) environmental dependence
in the Coma Supercluster relying purely on SDSS data, regardless of its well-known 
incompleteness at the bright  end (Blanton et al. 2005a,b,c), we decided to publish our measurements derived in
the past six years, with the hope that they might contribute to improving our understanding of AGNs.\\
There is a clear need to establish the frequency of AGNs of various types, in various environments, 
locally and in a cosmological perspective, to improve our understanding of galaxy assembly.
Nevertheless, even after the advent of SDSS, which has permitted many extensive studies of galaxies,
the processes involved remain unclear.
Hao et al. (2005) estimated that 4-10\% of all galaxies in the SDSS harbor an AGN;
Kauffmann et al. (2003) estimated that up to 80\% of galaxies more massive than $10^{11} M_\odot$
host a supermassive black hole, either dormant or active.
The role of the environment in triggering or inhibiting nuclear activity remains disputed. 
Kauffmann et al. (2004), analyzing $\sim 122000$ SDSS spectra, found a decrease in the fraction of powerful AGNs 
(with high [OIII] luminosity, i.e. dominated by Seyferts) with increasing galaxy environment density, mainly because the 
most powerful AGNs are hosted by late-type galaxies, which are the least likely to exist in dense environments.
Popesso \& Biviano (2006) found that the AGN fraction decreases with the increasing velocity dispersion 
of galaxies in groups and clusters,
being higher in dense, low-dispersion groups, in contrast to the findings of
Shen et al. (2007) and Miller et al. (2003) who found that the frequency of low activity AGNs 
does not correlate with environment. \\
The classification of AGNs based on optical spectra has been traditionally performed using the BPT diagnostic diagram of 
Baldwin et al. (1981), which adopts the measurement of at least four spectral lines: $\rm H\beta$, O[III], $\rm H\alpha$, 
and N[II].
The ratio $\rm H\alpha$/N[II] (where $\rm H\alpha$ must be corrected for any underlying stellar absorption, 
as stressed by Ho et al. 1997)
differentiates AGNs from HII region-like nuclei, while the ratio $\rm H\beta$/O[III] 
allows us to separate the strong AGNs from the weaker LINERs.
More recently Cid Fernandes et al. (2010, 2011) introduced a two-line 
diagnostic diagram, named WHAN, based on the $\rm H\alpha$/N[II] ratio combined with 
the strength of the $\rm H\alpha$ line (corrected for underlying stellar absorption) to discriminate both
strong and weak AGNs, which are both supposed to be ionized by active black holes, from "fake AGNs",
dubbed "retired galaxies", whose ionization mechanism is provided by their old stellar population. 
As argued by several authors (e.g. Trinchieri \& di Serego Alighieri 1991, Binette et al. 1994, Macchetto et al. 1996, Stasi{\'n}ska et al. 2008, Sarzi et al. 2010, 
Capetti \& Baldi 2011), hot evolved stars, such as post-asymptotic giant branch stars and white
dwarfs, can produce a substantial diffuse field of ionizing photons: it has been shown that they can produce emission lines with ratios
mimicking those of AGNs and - in particular - of LINERs.  \\
An equivalent width (hereafter EW) of $\rm H\alpha$ of at least 3 \AA ~is ultimately 
required for a nucleus to be considered 
ionized by a central black hole (Cid Fernandes et al. 2010, 2011).
This quantitative threshold of course might be questioned.
Among the galaxies in our sample, we might consider for example the elliptical NGC3862 (CGCG 97-127), 
which is the second brightest galaxy of A1367 and was classified as an AGN by
V{\'e}ron-Cetty \& V{\'e}ron (2006). It harbors the radio galaxy 3C-264 (Gavazzi et al. 1981)
and an optical jet (Crane et al. 1993), leaving little doubt that it contains a central monster.
Its optical spectrum (taken either at Loiano or from SDSS)
displays a strong NII emission line (of EW 4.81 \AA) and a weak $\rm H\alpha$ EW of 1.51 \AA. After adding 1.3 \AA~
to the underlying absorption, the galaxy just misses the  $\rm H\alpha>3~\AA$ threshold, 
thereby fails to be classified as a LIN but not as a RET according to the criteria 
of Cid Fernandes et al. (2010, 2011)\footnote{
Similar examples (see Section \ref{results}) convinced us to lower the threshold for LIN 
from 3 \AA~ to 1.5 \AA~ in the present analysis.}.\\
The Coma Supercluster, on which this paper is focused,
is an ideal laboratory for initiating a complete census of AGNs at $z=0$ because 
problems related to the incompleteness of SDSS are less severe here than elsewhere.
Even its largest sized galaxies have diameters $<2-3$ arcmin, making the "shredding" problem much less
severe than for other nearby environments (e.g. Virgo). The brightest galaxies have been well-studied
and even the faintest galaxies belonging to the Coma Supercluster
have sizes in excess of 10 arcsec, allowing their classification by visual inspection
on SDSS plates. Moreover, at the distance of 100 Mpc in the Coma Supercluster, the three arcsec fibers adopted by SDSS
provide truly "nuclear" spectra. 
Studying the Coma Supercluster using both the integrated and nuclear properties of galaxies at $z=0$
derived from SDSS data, allows us to compare galaxies properties for 
a variety of environmental conditions, from the center of two rich clusters 
to the sparse filamentary regions where the local galaxy density is $\sim$ 100 times lower,
without the biases caused by studying galaxies at different distances.
It provides a complementary view to
the many statistical analyses that have been made using SDSS data
(e.g. Kauffmann et al. 2004).
Combined with studies of galaxy evolution at higher redshift
it will shed light on the mechanisms and processes that contribute to the evolution of galaxies 
in the cosmological context. \\
Owing to the two-line WHAN diagnostics, the red-channel spectra taken at Loiano can contribute significantly to the census of AGNs 
by filling the residual incompleteness of the SDSS spectral database for luminous galaxies caused by 
shredding and fiber conflict 
(Blanton et al. 2005a,b,c).\\
The layout of this paper is as follows.
In Sect. \ref{The data}, we illustrate the observations taken at the Loiano observatory 
and the data reduction procedures. The nuclear spectra are given and classified in Sect. \ref{results}.
The census of AGNs of the various types in the Coma Supercluster is carried out in Sect. \ref{AGN COMA}. 
Our results are compared with Mahajan et al. (2010) and discussed in Sect. \ref{Discussion}.
The standard cosmology is assumed, with $Ho$=73 $\rm km~sec^{-1}~Mpc^{-1}$.

\section{The sample}
\label{sample}

\begin{figure*}
\begin{center}
\includegraphics[width=17cm]{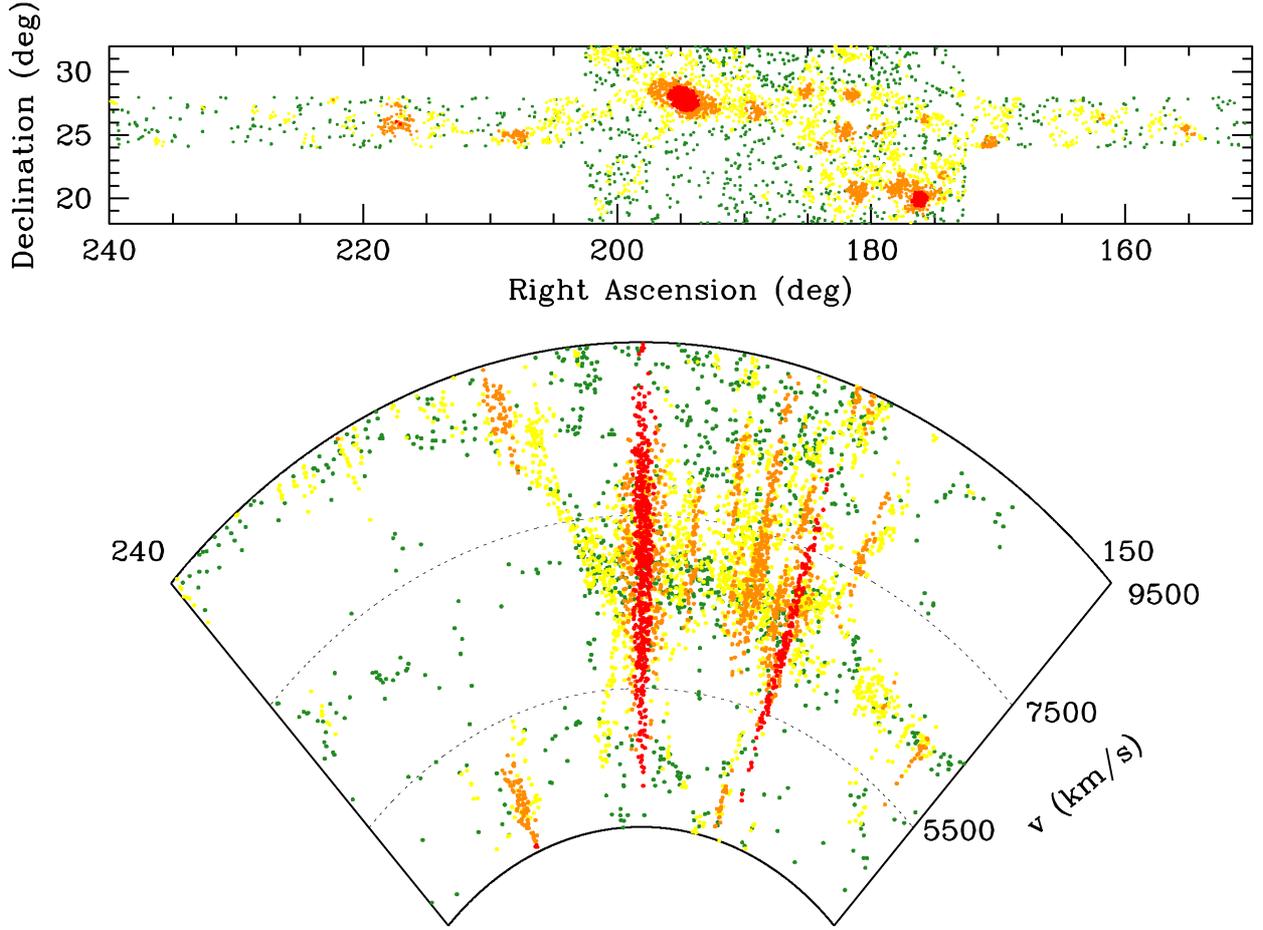}
\caption{({\bf Top panel}) The Coma Supercluster region studied in this work in celestial coordinates and as
a wedge-diagram ({\bf Bottom panel}).
Symbols are color coded according to the density parameter (see Section \ref{AGN COMA}): UL: $\delta_{1,1000}\leq 0$ are green;  L: $0 < \delta_{1,1000} \leq 4$ are yellow;
 H: $4 < \delta_{1,1000} \leq 20$ are orange; UH: $\delta_{1,1000} > 20$ are red.}
\label{wedge}
\end{center}
\end{figure*}

\begin{figure}												          
\includegraphics[width=8.6cm]{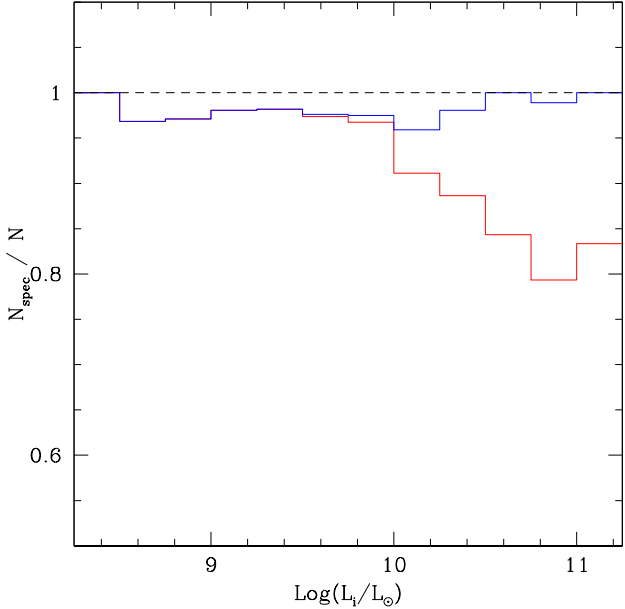}								          
\caption{The completeness of the "COMA sample", 
consisting of galaxies taken from SDSS, targets taken from		          
CGCG, NED, and ALFALFA, and objects with spectra taken at Loiano (blue), is 98\%	          
(only 112 galaxies remaining without nuclear spectra).							          
Galaxies selected from the SDSS alone (red), e.g. used by Mahajan et al. (2010) are incomplete by $\sim 20 \%$    
in the highest luminosity bins, where most AGNs are expected.}									          
\label{compl}												          
\end{figure}

This paper is based on two samples:\\
i) A miscellaneous set of 283 relatively bright galaxies, not complete by any means, 
for which we took long-slit, red-channel nuclear spectra of R $\sim$ 
1000, with the Loiano telescope. We refer to this as the "LOI sample" hereafter.\\
ii) A complete set of 5027 galaxies (hereafter the "COMA sample") selected in the region 
$11.5^h<RA<13.5^h;+18^o<Dec<32^o$ (see GoldMine, Gavazzi et al. 2003)
that traditionally describes the Coma Supercluster and was extensively studied  by Gavazzi et al. (2010), 
combined with an extended region defined by $10^h<RA<16^h$ and $+24^o<Dec<28^o$ (see Fig. \ref{wedge}).
The second area is suitable for filling an LST interval corresponding 
to the duration of an observing night in spring
and has also been covered with HI observations by ALFALFA (Giovanelli et al. 2005), whose HI selected 
targets will shortly be publicly available.\\
Following the method of Gavazzi et al. (2010),
we searched in this area of the sky ($11.5^h<RA<13.5^h;+18^o<Dec<32^o$ combined with $10^h<RA<16^h;+24^o<Dec<28^o$) 
the SDSS DR7 spectroscopic database (Abazajian et al. 2009) for all galaxies 
with $r\leq$17.77 in the redshift interval $3900<cz<9500~{\rm km~s^{-1}}$. 
We obtained 4790 targets. For each, we extracted the coordinates, 
$u,g,r,i,z$ Petrosian magnitudes (AB system), and spectroscopic information, 
including the principal line intensities and the redshift. 
The morphological classification of all galaxies was performed by individual visual inspection of
SDSS color images. \\
To fill-in the incompleteness of SDSS for luminous galaxies due to shredding and fiber conflict 
(Blanton et al. 2005a,b,c), we added 133 CGCG (Zwicky et al. 1961-1968)
galaxies with known redshifts from NED that were not included in the SDSS spectral database (see Fig. \ref{compl}).  
For these objects, we also measured the $u,g,r,i,z$ Petrosian magnitudes using the SDSS DR7 navigation tool,
which provides accurate magnitudes.\\
Additional galaxies that could not be found in the SDSS spectroscopic catalog were searched for using NED.
For these targets, we again evaluated 
the $u,g,r,i,z$ magnitudes using the SDSS navigator tool and among them we selected 76 objects meeting 
the condition $r\leq17.77$, which matches the selection criterion of the SDSS spectral catalog (Strauss et al. 2002).
We repeated a similar search in the ALFALFA database (Haynes, private communication) for the region $10^h<RA<16^h$ and 
$+24^o<Dec<28^o$,
and found 28 additional HI-selected systems with $r\leq$17.77, not included in the SDSS spectral database. 
In total, our sample consists of 5027 galaxies: 4790 from SDSS and 237 from other sources.\\
The 4785 out of 4790 galaxies selected from SDSS have a spectrum available from the SDSS database (five remaining objects
have partly corrupted spectra); another 130 were observed at Loiano, taken from sample i), and the remaining 112 
do not yet have available spectra. In total, the "COMA sample" reaches a spectral completeness of 98\%.

\section{Observations and data reduction}
\label{The data}

Optical spectra of the nuclei of 283 galaxies in sample i) were obtained during 14 runs from 2005 to 2011 (see Table \ref{Table1}) 
using the Bologna Faint Object Spectrograph and Camera
(BFOSC) attached to the 152cm F/8 Cassini Telescope located
in Loiano, belonging to the Observatory of Bologna.
These consist of long-slit spectra taken through a two-arcsec slit, combined with an intermediate-resolution grism (R $\sim$ 1000)
covering the 6100 - 8200 \AA ~portion of the red-channel and containing $\rm H\alpha$, [NII], and [SII] spectral lines.
BFOSC is equipped with a EEV LN/1300-EB/1 CCD detector of 1300x1340 pixels, reaching 90\% QE near 5500 \AA. 
For the spatial scale of 0.58 arcsec/pixel, the useful field is of 12.6$\times$13 square arcmin.
The grism dispersion of 8.8 nm/mm results in spectra with 1.6 \AA/pix.
Exposures of 3-5 minutes were repeated typically three times (to help remove the cosmic ray hits), up to six times.
The slit was generally set in the E-W direction, except when aligned along the galaxy major axis,
or along the direction connecting two nearby objects taken simultaneously in the slit. The wavelength calibration  was secured 
by means of frequent exposures of a He-Ar hollow-cathode lamp.
The flux calibration and the spectrograph response were obtained by daily exposures of the star Feige-34. 
The typical seeing conditions at Loiano ranged from 1.5 to 2.5 arcsec.\\
\begin{table}[ht!]
\caption{Log of the observations at Loiano. Two new-moon periods of four nights each were allocated to the present 
project per year. We list information for the useful 
nights. The total number of spectra (305) includes 22 of objects that were repeatedly observed during several runs. 
In these cases, Table \ref{Table2} describes the combined spectra.}
\centering
{\footnotesize \begin{tabular}{llll}
\hline
\hline
Year & Feb & Mar &   Nspec \\
2005  &   9,10,11     &     9,11,12   &  45  \\      
2006  &   26          &  23           &  24  \\  
2007  &   13,15,16    &  17,18	      &  29  \\  
2008  &   5	      &   3           &  10  \\
2009  &   18,19,21    &   26,27	      &  54  \\
2010  &   8 	      &    -          &  10  \\
2011  &   7,8,9,10    &   6,7,8       &  133 \\
\hline
\end{tabular}
}
\label{Table1}
\end{table}
The reduction of the spectra was based on the IRAF 
STSDAS\footnote{IRAF is the Image Analysis and
Reduction Facility made available to the astronomical community by the
National Optical Astronomy Observatories, which are operated by AURA,
Inc., under contract with the U.S. National Science Foundation. STSDAS
is distributed by the Space Telescope Science Institute, which is
operated by the Association of Universities for Research in Astronomy
(AURA), Inc., under NASA contract NAS 5--26555.} 
reduction packages. 
The wavelength calibration was performed on the lamp exposures using $identify$$-reidentify$$-fitcoo$ and transferred to the scientific
exposures using 
$transform$. After checking the wavelength calibration using the principal sky lines, these were removed using $background$.
One-dimensional (1d) spectra of the nuclei were extracted from the two-dimensional (2d) images using 
$apsum$ in apertures of 5.8 arcsec. 
After flux calibration, the 1d spectra were
normalized to the intensity of the continuum under $\rm H\alpha$, transformed to rest-frame velocity,
and measured with $splot$
to obtain the EW of the $\rm H\alpha$ and $\lambda$ 6584 [NII] lines.

\section{Spectral classification}
\label{results}

The classification of the nuclear activity based on optical spectra was performed according to identical rules
for sample i) LOI and ii) COMA (except for Seyfert galaxies; see below in this section).
The classification rule is
based on the WHAN diagram introduced by Cid Fernandes et al. (2010, 2011), which is based on the strength of
the $\rm H\alpha$ line corrected for underlying stellar absorption (Ho et al. 1997) 
and the ratio of [NII] to $\rm H\alpha_{corr}$ 
(EW) to discriminate between different levels of nuclear activity.
As for the underlying absorption at $\rm H\alpha$, Decarli et al. (2007) 
adopted an average correction of $1.7~\AA$.
Owing to the limitations of the SDSS spectral database, this quantity is here determined 
again to greater accuracy 
using 881 SDSS spectra of passive galaxies in the Coma Supercluster (see Table \ref{Table3}).
The mean underlying stellar absorption at $\rm H\alpha$  present in all passive galaxy spectra amounts to 1.3 $\AA$,
irrespective of luminosity, below or above $\rm Log (L_i/L_{\odot})=10$)\footnote{
For star-forming galaxies, the direct determination of absorption 
is hampered by the presence of $\rm H\alpha$  and [NII] in emission. 
However, for bright spirals ($\rm Log (L_i/L_{\odot})>10$, often dominated by bulges)
the presence of underlying absorption at $\rm H\alpha$ is sometimes detected 
(although difficult to quantify) in the SDSS spectra.
In contrast, for faint late-types ($\rm Log (L_i/L_{\odot})\leq 10$) the absorption component 
at $\rm H\alpha$ is generally not detected, being overtaken  
by the emission component, which amounts on average to 30 \AA~ (and exceeds 5 \AA~ for 93\% of them), 
making the 1.3 ~\AA~ correction negligible.}. 
Thus we adopt 1.3 $\AA$ for the underlying continuum absorption at $\rm H\alpha$, 
i.e. $\rm H\alpha_{corr} = H\alpha - 1.3 ~\AA$. \\
\begin{figure*}
\begin{centering}
\includegraphics[width=15cm]{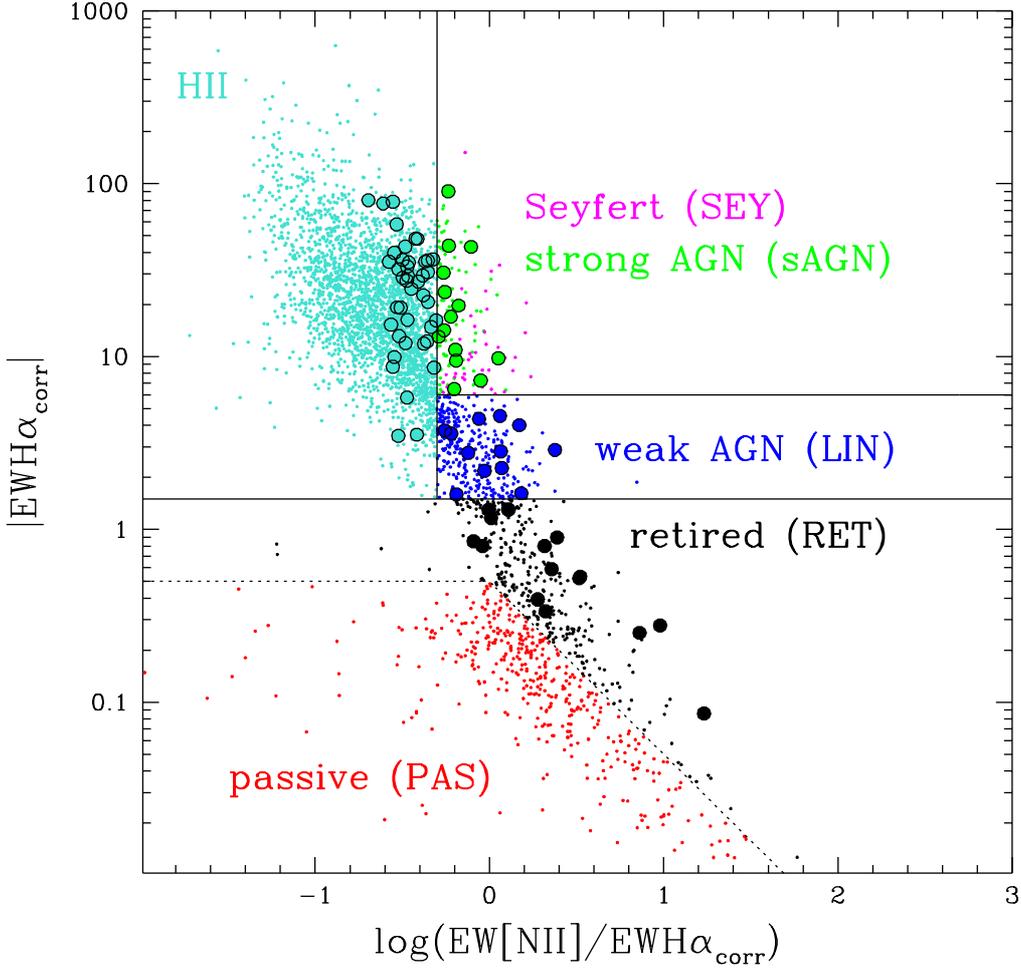}
\caption{The WHAN diagnostic diagram. The equivalent width of $\rm H\alpha$ is corrected by 1.3 \AA~ 
for underlying absorption. Small symbols refer to the COMA sample, large symbols to the LOI sample. Seyfert1 galaxies, 
identified visually in the Loiano spectra, are given as triangles.}
\label{whan}
\end{centering}
\end{figure*}
Fig. \ref{whan} shows the WHAN diagram with the adopted spectral classification thresholds. 
The vertical line shows the separation between HII region-like nuclei and AGNs.
This threshold is set at $\log(EW {\rm [NII]}/EW {\rm H\alpha_{corr}}) = -0.30$ in agreement with Capetti \& Baldi (2011), 
as a compromise between the value of -0.22 adopted by
Baldwin et al. (1981) and Decarli et al. (2007), and the value of -0.40 adopted by Cid Fernandes 
et al. (2010, 2011)\footnote{
With this choice, two galaxies (NGC3773 and CGCG136-042) classified as HII regions in NED
are inconsistently classified as LINER by ourselves because they have $\rm -0.3 < \log(EW {\rm [NII]}/EW {\rm H\alpha_{corr}}) < -0.22$,
but  another two (NGC2537 and CGCG71068), classified as HII regions by both NED and us, would have been inconsistently classified as LINERS
by Cid Fernandes et al. (2010, 2011).}. \\
In Fig. \ref{whan}, several $EW\rm H\alpha_{corr}$ thresholds are indicated. These are used to 
discriminate between the various 
levels of nuclear activity, as recommended by Cid Fernandes et al. (2010, 2011)
with  one difference: the separation between both HII regions and weak AGNs (LIN) from "fake ones" (RET) is
set to $|EW\rm H\alpha_{corr}| = 1.5~\AA$ instead of 3~\AA \footnote{
This choice is adopted because well-established LINERS 
e.g. VCC1110 (V{\'e}ron-Cetty \& V{\'e}ron (2006)), VCC2070 (Panessa \& Bassani 2002), 
NGC 2841 (V{\'e}ron-Cetty \& V{\'e}ron (2006)), 
NGC 6166 (containing the radio galaxy 3C-338, Gavazzi \& Perola 1980), CGCG 97-127 (V{\'e}ron-Cetty \& V{\'e}ron (2006), 
containing the radio galaxy 3C-264, Gavazzi et al. (1981),
with optical jet Crane et al.1993), which would otherwise have been classified as RET by Cid Fernandes et al. (2010, 2011),
are consistently classed among LIN.
Even after weakening the criterion for LIN to $|EW\rm H\alpha_{corr}| > 1.5~\AA$, some inconsistencies remain
between our classification and NED:
NGC2685 (classified as Sey2 by V{\'e}ron-Cetty \& V{\'e}ron 2006) and VCC1615 (classified as LIN by V{\'e}ron-Cetty \& V{\'e}ron 2006) 
are classified as RET in this work.}.
Summarizing, we define six independent classes of nuclear activity:
\begin{itemize}  
\item "HII region-like" nuclei are characterized by intense star-formation activity: their 
spectra display both $\rm H\alpha$ and [NII] in emission,
with $|EW\rm H\alpha_{corr}| > 1.5~\AA$ and the ratio $\log(EW {\rm [NII]}/EW {\rm H\alpha_{corr}})<-0.3$;
\item strong Active Galactic Nuclei (sAGN) contain a powerful active nucleus (possibly 
triggered by a supermassive black hole): their 
spectra have $|EW\rm H\alpha_{corr}| > 6~\AA$ and $\log(EW {\rm [NII]}/EW {\rm H\alpha_{corr}})>-0.3$;
\item weak AGNs (LIN) show the signature of a less active nucleus than sAGN: their spectra
have $1.5 < |EW\rm H\alpha_{corr}| < 6~\AA$ and $\log(EW {\rm [NII]}/EW {\rm H\alpha_{corr}})>-0.3$;
\item fake AGNs (RET) are thought to be "retired nuclei" 
whose ionization mechanism is provided by their old stellar population in a way that can mimic 
the processes of AGNs and 
- in particular - of LINERs: their spectra 
have either $\rm H\alpha$ in moderate absorption ($0.5  < |EW\rm H\alpha_{corr}| < 1.5~\AA$) or both
$\rm H\alpha$ in absorption ($|EW\rm H\alpha_{corr}| < 0.5~\AA$) and [NII] in emission ($|EW[NII]| > 0.5~\AA$);
\item passive nuclei (PAS) are typically red, contain exclusively old stars, with no star formation activity: their spectra 
show both $\rm H\alpha$ and [NII] in absorption ($|EW\rm H\alpha_{corr}| < 0.5~\AA$ and $|EW[NII]| < 0.5~\AA$).
\end{itemize} 
The classification of Seyfert galaxies follows different rules for the "LOI sample" and SDSS spectra. 
For SDSS spectra, we use $\rm EW[OIII] / EWH\beta > 3$ to identify Seyfert (SEY) among sAGN. 
In the "LOI sample", the absence of blue-channel spectra prevents us from using the BPT diagnostic diagram, hence 
Seyfert1 galaxies (SY1, represented with pink triangles
in Fig. \ref{whan}) are classified by visual
inspection of individual spectra for the presence of broad permitted lines.
Tables \ref{Table2} and \ref{Table3} summarize the classification of spectra, respectively, 
in the "LOI" and "COMA" samples.

\begin{figure*}
\begin{center}
\includegraphics[width=11cm]{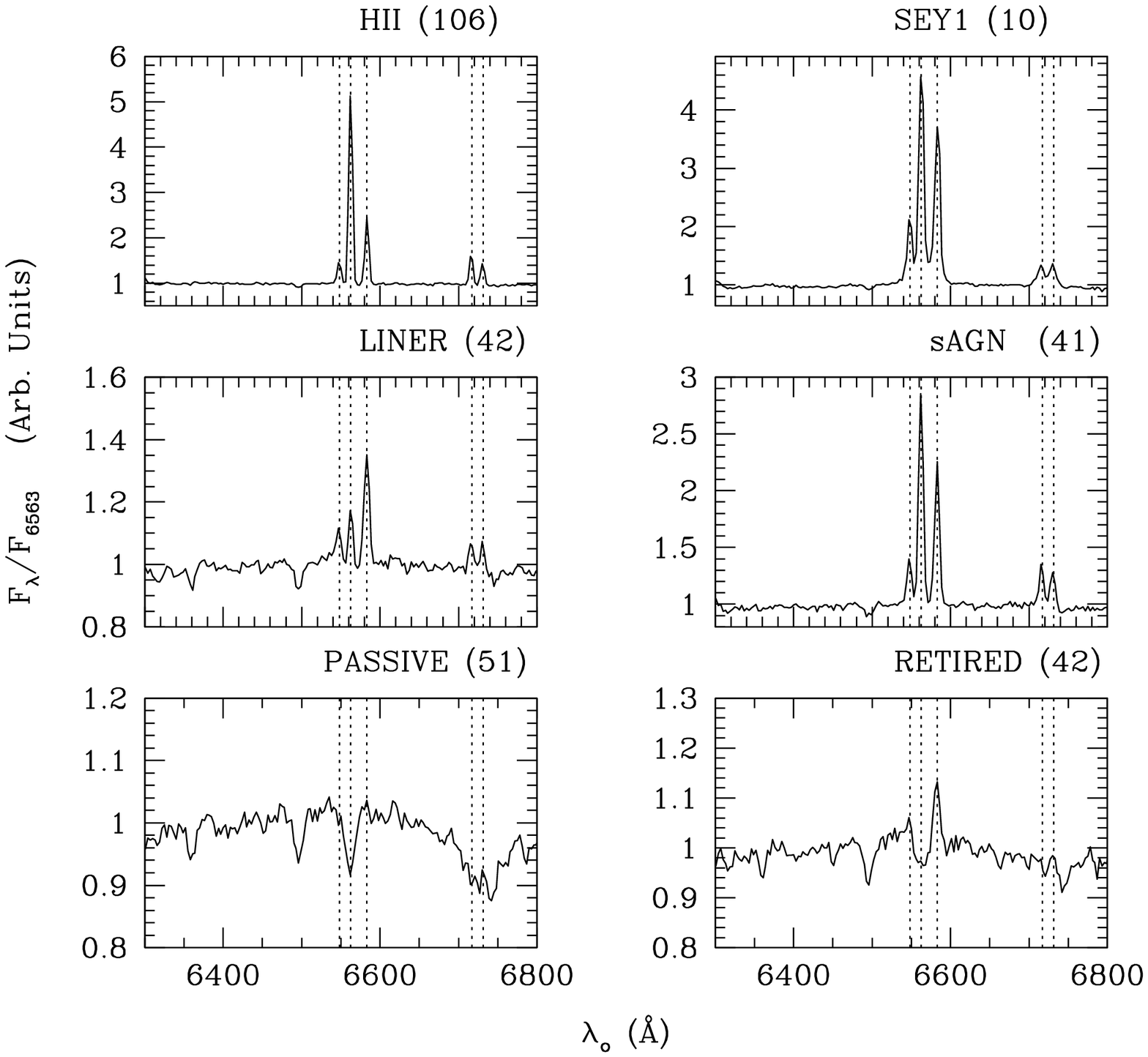}
\includegraphics[width=11cm]{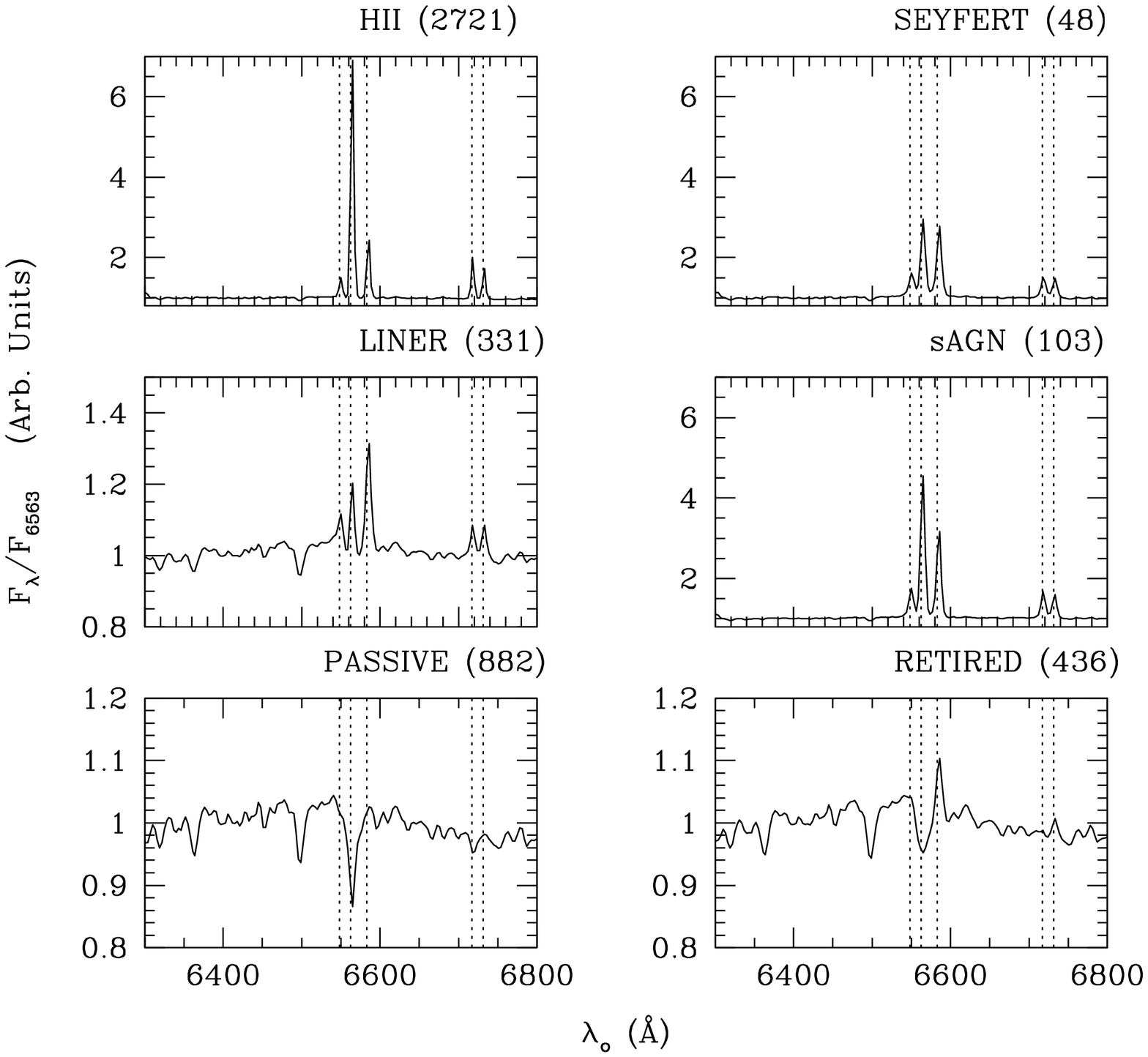}
\caption{Template spectra obtained by stacking individual normalized rest-frame spectra (whose number is given in parenthesis) 
in six bins of nuclear classification.
The vertical dashed lines mark the position of [NII] $\lambda$ 6548, $\rm H\alpha$ $\lambda$ 
6563, [NII] $\lambda$ 6584, and the [SII] doublet $\lambda$ 6718 and $\lambda$ 6733. "LOI sample ({\bf top panel});
"COMA sample" ({\bf bottom panel}).}
\label{templ}
\end{center}
\end{figure*}

\begin{figure}[!h]
\includegraphics[width=9cm]{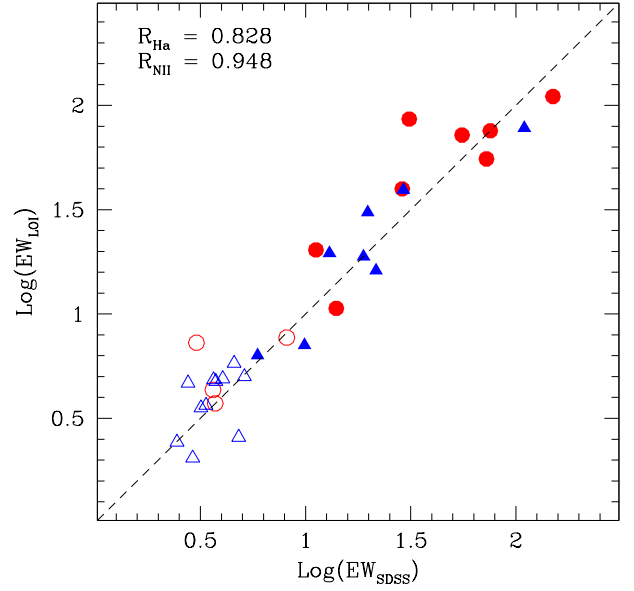}
\caption{The equivalent width of $\rm H\alpha$ (red dots) and $\lambda$ 6584 [NII] (blue triangles) for 15 galaxies in the Coma Supercluster that
were measured both at Loiano and by SDSS. The correlation coefficients are given. The dashed line gives the 1:1 relation.  Empty symbols refer to
galaxies with luminosity $\rm Log (L_i/L_{\odot})>10.5$.}
\label{corr}
\end{figure}

Template spectra of the six spectral classes, obtained by stacking N spectra taken at Loiano, 
are given in Fig. \ref{templ} (LOI top panel).
The 4785 FITS spectra available from SDSS were downloaded, shifted to their rest-frame using $IRAF dopcor$, 
and normalized to the flux intensity near $\rm H\alpha$.
The rms of their continuum were measured between 6230 \AA~ and 6490 \AA~ 
(avoiding bright galaxy spectral lines and sky residuals), providing 
an estimate of the signal-to-noise ratio (hereafter S/N). 
Spectra with S/N$>$10  were stacked  in each spectral class and are shown in  Fig. \ref{templ} (COMA bottom panel).\\
The parameters of the 283 individual spectra for the "LOI sample" are listed in Table \ref{Table4} as follows:\\
Column 1: Galaxy name;\\
Column 2 and 3: J2000 celestial coordinates;\\
Column 4: redshift from NED;\\
Column 5 and 6: measured EW of $\rm H\alpha$ and $\lambda$ 6584 [NII] lines from the present work
(negative values represent emission);\\
Column 7: the nuclear activity classification in six bins;\\
Column 8: other spectra available from either SDSS or NED (\emph{SDSS-corr} means that the SDSS spectrum is partly corrupted);\\
Column 9: independent classification of the nuclear activity from NED.\\
As anticipated in Sect. \ref{sample}, the 283 "LOI sample" galaxies do not form a 
complete sample in any sense. 
Among them, 41 belong to the Virgo cluster, which are not considered further here.
Not all of the 177 galaxies belonging to the Coma Supercluster have 
new spectroscopic measurements: several have also a nuclear spectrum
(acquired using three arcsec fibers) from the DR7 of SDSS and many others have nuclear spectra available from NED. 
Nevertheless, there are 130 Coma galaxies in the "LOI sample" (see Table \ref{Table4}) that do not 
have a spectrum in SDSS, hence these data contribute 
toward the completion of the "COMA sample" (see Section \ref{AGN COMA}).
For 15 Coma galaxies taken at Loiano before the DR7 of SDSS was issued, Fig. \ref{corr} displays the two independent
measurements of the equivalent width of $\rm H\alpha$ and $\lambda$ 6584 [NII], which do not appear 
to be affected by systematic errors\footnote{The extraction of the Loiano spectra
was carried out in apertures whose area (11.6 $\rm arcsec^2$) is almost twice as large as that of the SDSS fibers 
(7.1 $\rm arcsec^2$). This was done on purpose to compensate for the Loiano spectra 
being taken for the brightest galaxies, which are a factor of two, on average, larger than those taken by SDSS. Fig. \ref{corr} shows no systematic
effects for bright and faint objects.}.
\begin{table}[h!]
\caption{Line measurements performed on stacked Loiano spectra of N objects in six spectral classes in the "LOI sample" 
(negative = emission).}
\centering
{\footnotesize \begin{tabular}{lccc}
\hline
\hline
Class &	N  &$EW \rm H\alpha$    &	$EW[NII2]$  \\
\hline
HII      &  106	&    -25.4  &    -9    \\
SY1      &   10 &    -30  &    -24     \\
sAGN      &   41	&    -13.6 &    -9.1   \\
LIN      &   42	&    -1.7 &    -3.0    \\
RET      &   12 &     1.0 &    -1.1    \\
PAS      &   42	&     1.6 &    0.0     \\
\hline
\end{tabular}
}
\label{Table2}
\end{table}
\begin{table}[!h]
\caption{Line statistics in the "COMA sample" containing N objects in each spectral class
(negative = emission).}
\centering
{\footnotesize \begin{tabular}{lccc}
\hline
\hline
Class &	N  &	  $EW \rm H\alpha$    &	$EW[NII2]$   \\
\hline
HII   & 2721	&    -28.7   &  -5.9  \\
SEY   &   48	&    -13.4   &  -12.6   \\
sAGN   &  103	&    -22.8   &  -14.8  \\
LIN   &  331	&    -1.7    &  -2.5  \\
RET   &  436	&     1.0    &  -1.0  \\
PAS   &  881	&     1.3    &  -0.2  \\
\hline
\end{tabular}
}
\label{Table3}
\end{table}

\section{AGNs in the Coma Supercluster}
\label{AGN COMA}
 \begin{figure*}
 \begin{centering}
\includegraphics[width=11cm]{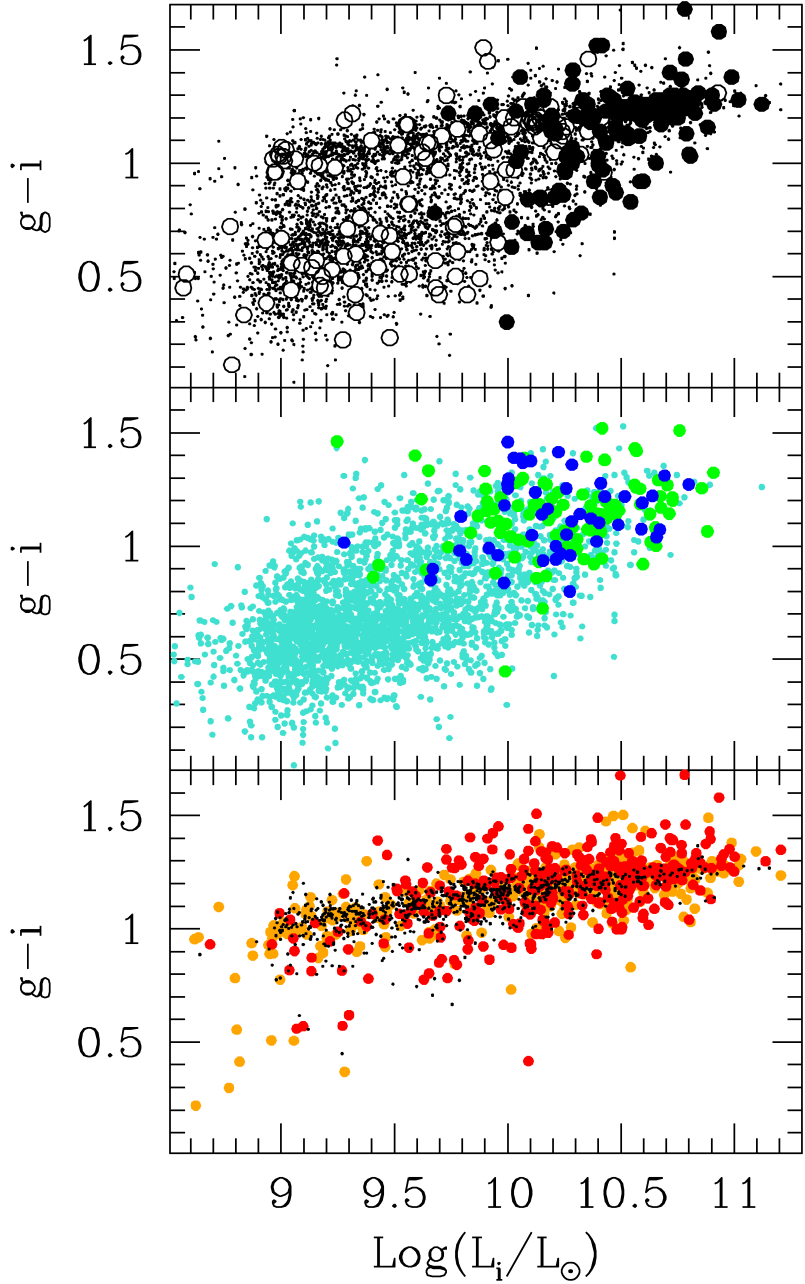}
\caption{The color (g-i) vs $i$-luminosity diagram of galaxies studied in this work. 
{\bf Top panel}: small black dots represent the 4900 galaxies from SDSS;
large filled symbols mark the 130 galaxies for which a nuclear spectrum was taken at Loiano, 
and large open symbols mark the 112 galaxies yet without spectra.
{\bf Middle panel}: HII regions (turquoise), SEY (blue), and sAGN (green) form the blue sequence.
{\bf Bottom panel}: LIN (red), RET (orange), and PAS (black) form the red sequence.}
\label{colmag}
\end{centering}
\end{figure*}
 \begin{figure}
\includegraphics[width=8cm]{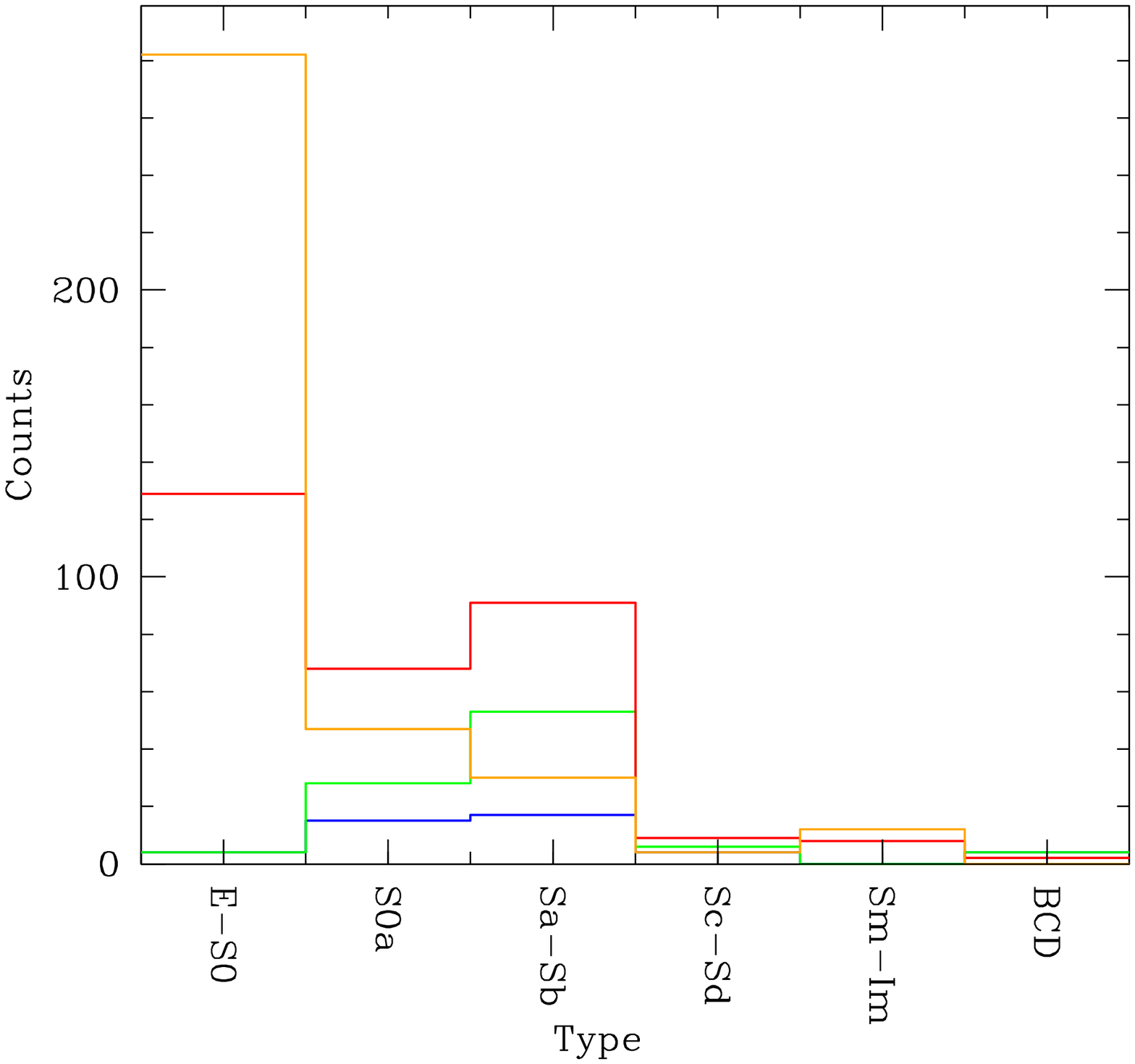}
\includegraphics[width=8cm]{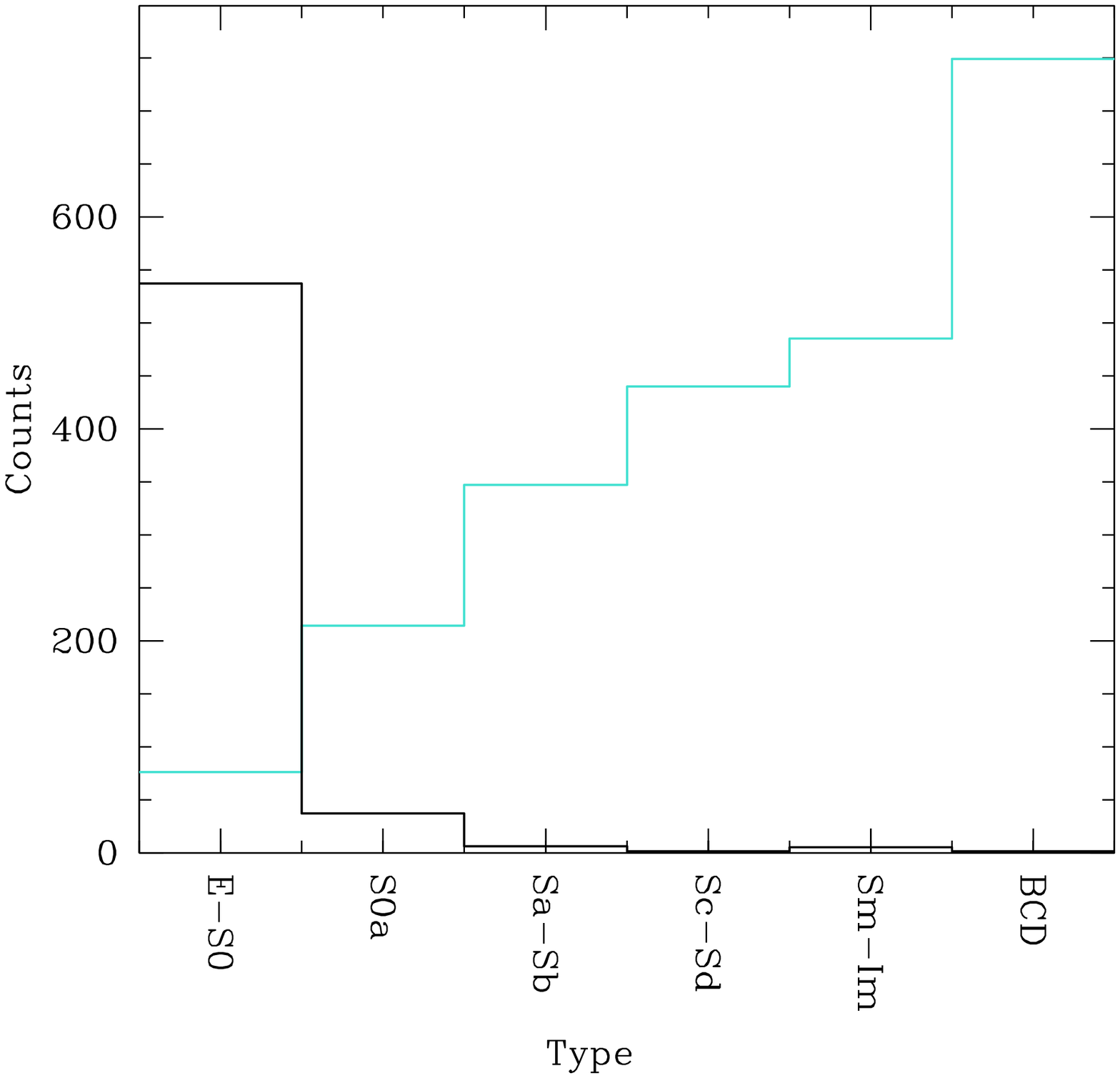}
\caption{The morphology distribution of: ({\bf top panel}) SEY (Blue), LIN (red), sAGN (green), RET (orange);
({\bf bottom panel}) HII (turquoise), PAS (black). }
\label{morf}
\end{figure}
\begin{figure}
\includegraphics[width=8cm]{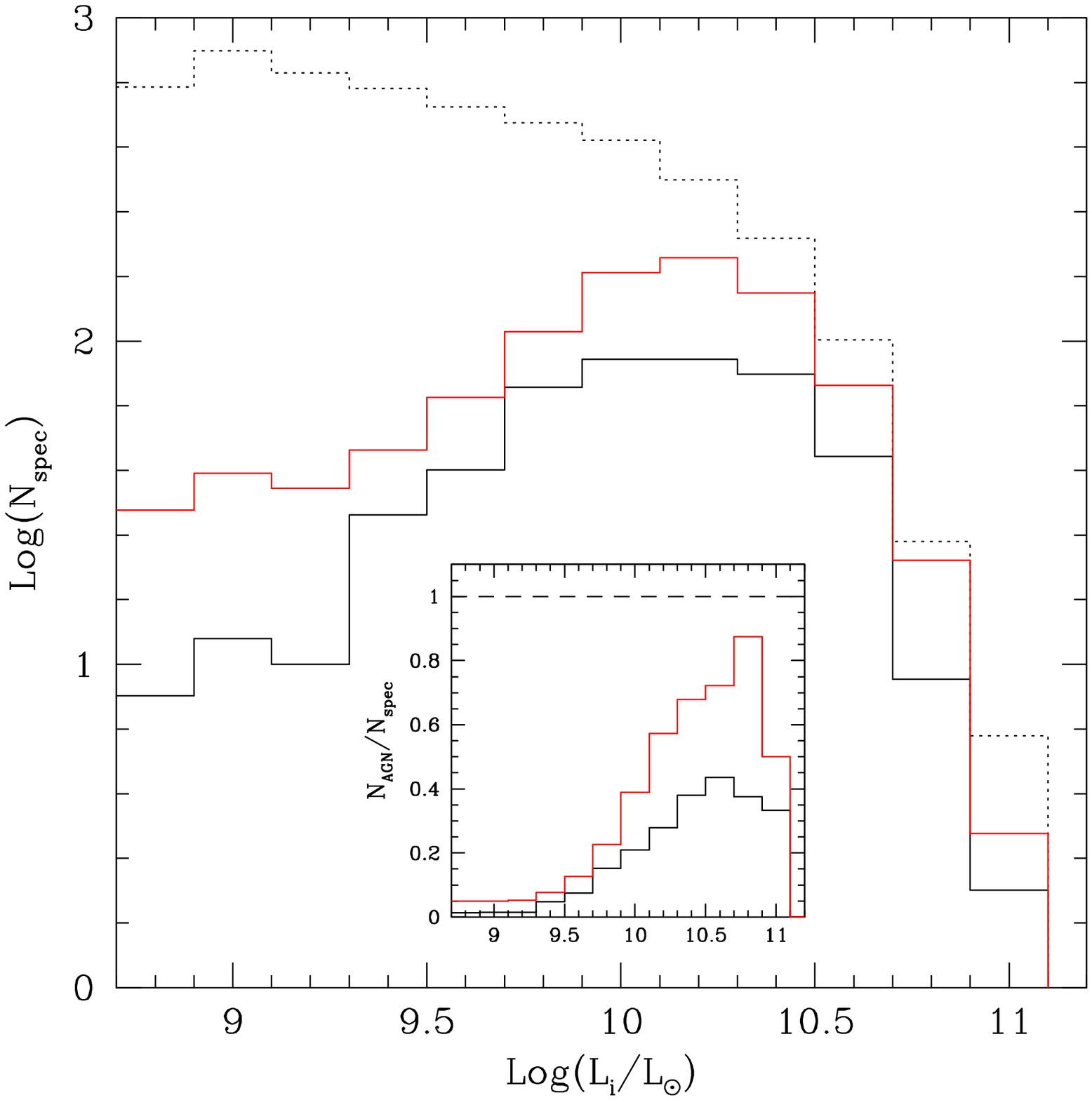}
\caption{The luminosity distribution of AGNs with (red) and without (solid black) RET  and of all galaxies (dotted).
The inset shows the ratio of AGNs to all galaxies with (red) and without (solid black) RET.}
\label{lum}
\end{figure}
The data obtained in the previous section are used to help perform a census of AGNs in the Coma Supercluster
region. \\
Following an identical procedure to Gavazzi et al. (2010), the local number density $\rho$ around each galaxy was 
computed within a cylinder 
of 1 $h^{-1}~\rm Mpc$ radius and 1000 $\rm km~s^{-1}$  half-length.
We divide the sample into four overdensity bins, chosen in order to highlight physically different
environments of increasing level of aggregation:
the Ultra-Low density bin (UL: $\delta_{1,1000}\leq 0$) describing the underlying cosmic web;
the Low density bin (L: $0 < \delta_{1,1000} \leq 4$) comprising the filaments in the Great Wall and the
loose groups; the High density bin (H: $4 < \delta_{1,1000} \leq 20$) including
the cluster outskirts and the significant groups; and the Ultra-High density bin (UH: $\delta_{1,1000} > 20$) 
corresponding to the cores of rich clusters.\\
Fig. \ref{wedge} shows the celestial distribution and the wedge diagram for the 5027 galaxies in the "COMA sample", 
coded according to the four density bins.\\
Fig. \ref{colmag} (top panel) gives the $g-i$ color vs. $i$-band luminosity diagram for the studied galaxies,
highlighting the distribution of galaxies without an SDSS spectrum and that the spectra taken at 
Loiano (filled dots) were obtained deliberately for bright objects with $\rm Log (L_i/L_{\odot})>10$, 
where the missing spectra from SDSS are more abundant and AGNs are more frequently found
(Decarli et al. 2007). Out of the 112 galaxies remaining with no spectra (empty circles), only 30 
are brighter than $\rm Log (L_i/L_{\odot})=10$. Among them, we expect to miss approximately 10 AGNs, given
the luminosity distribution of AGNs discussed below. \\ 
Before discussing the various color diagrams, we emphasize that the contribution of AGNs to the total $i$ luminosity is 
negligible, as the light measured in nuclear apertures of three arcsec diameter for the most luminous of all AGNs, i.e. the Seyferts, 
represents between 5\% and 15\% of the total light. \\  
Fig. \ref{colmag} (middle panel) shows the position of HII, SEY, and sAGN in the color-magnitude diagram,
clearly showing that AGNs of high activity belong to the high luminosity tail of the sequence composed of
star-forming galaxies (their slope is 0.34, and the dispersion is 0.20). 
Fig. \ref{colmag} (bottom panel) illustrates the position of PAS, RET, and LIN, showing that AGNs of low activity,
including fake-AGNs, belong to the high luminosity part of the "red sequence" (their slope is 0.17, and the dispersion is 0.10).
LIN follow a relation with a slightly higher slope (0.20) and dispersion (0.14) than
RET and PAS alone (slope = 0.16, and dispersion = 0.09), suggesting that they might belong to a transition class
between  early- and late-type galaxies.
From Fig. \ref{colmag} (middle and bottom panels), we can see that many luminous objects, mostly AGNs, have
redder colors than red-sequence galaxies of similar luminosity and in particular of PAS.
These galaxies are strongly affected by extinction from dust lanes. \\
We have measured the average $g-i$ for the whole galaxy in the three classes of AGNs of decreasing activity (SEY, sAGN, LIN)
and among RET. The resulting mean color for the four classes is $<g-i> \simeq 1.15 \pm 0.15$ mag, consistent with 
the average color of galaxies in the red sequence and of PAS galaxies.
This value is not found to depend on the local galaxy density.
The only class with $<g-i> \simeq 0.72$, which is significantly bluer 
than all other classes, is the one of HII-like nuclei.
Since $g-i$ color can be assumed as a proxy of the star formation in galaxies, this result seems to support 
the evolutionary scenario where the feedback from AGNs produces a significant quenching of the star formation 
in their host galaxies (e.g. Schawinski et al. 2009). 
However, our finding neither proves nor disproves 
this hypothesis because the same mean color might be found for PAS galaxies, which do not necessarly originate from 
a previous AGN phase, but simply from the fading of a nuclear starburst phase. \\                                    
The morphology distribution of AGNs, including RET, is given separately in Fig.
\ref{morf} (top panel). It appears that sAGN and SEY are 
mostly (40\%) associated with giant spiral galaxies (Sa-Sb) (and another 29\% to S0a).
LIN and RET are 78\% associated with early-type galaxies (E-S0a).\\
Among non-AGNs (Fig. \ref{morf} bottom panel), PAS are 94\% associated with E-S0, while HII regions
are 87\% associated with late-type galaxies (Sa-BCD).\\
The differential luminosity distribution of all AGNs (SEY+sAGN+LIN) (with and without RET) 
is given in Fig. \ref{lum}.
For all galaxies, the luminosity distribution follows a 
Schechter function (whose parameters can be found in Gavazzi et al. (2010). 
For the AGNs, the distribution is instead "Gaussian", with a peak near $\rm Log (L_i/L_{\odot})=10.2$.
Below this luminosity ($\rm Log (L_i/L_{\odot}) \leq 10.2$),
the fraction of Active Nuclei is 11\% with RET and 6\% without.
For $\rm Log (L_i/L_{\odot})>10.2$, it becomes 60\% with and 32\% without RET. 
RET alone represent a similar fraction of all bona-fide AGNs at any luminosity.\\

\begin{figure}
 \begin{center}
\includegraphics[width=9cm]{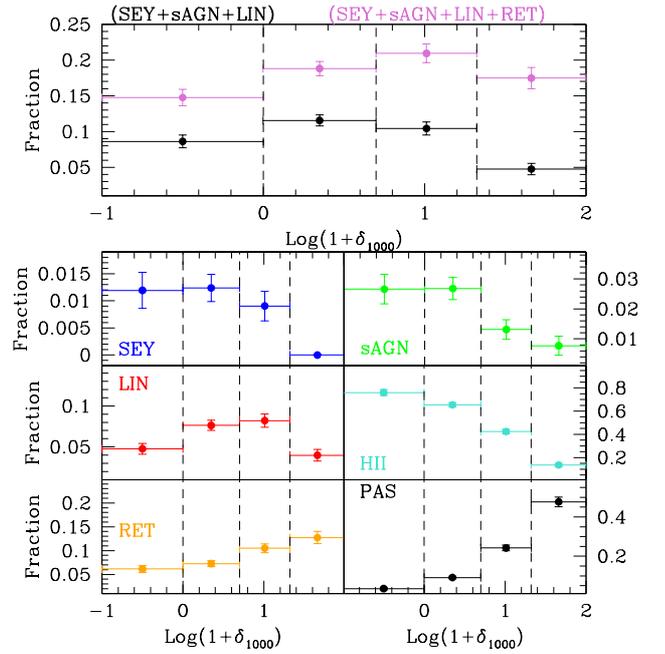}
\caption{The frequency distribution of SEY ({\bf top left}); sAGN ({\bf top right}); LIN ({\bf middle left}),
 HII ({\bf middle right}); RET ({\bf bottom left}); PAS ({\bf bottom right}) and all AGNs ({\bf top}) with RET (pink) and without RET (black)
 in 4 bins of local galaxy density. }
\label{dens}
\end{center}
\end{figure}

 When the frequency of galaxies in the various spectral classification bins (normalized to the total 
 number of surveyed galaxies) is plotted in bins of local galaxy density, it appears (see Fig. \ref{dens}) that:
 \begin{itemize}
 \item among AGNs, the percentage of SEY-sAGN-LIN does not change significantly from UL to H density but
 decreases by a factor of two in the densest UH environment
 (notice that LIN are as frequent in the lowest and highest density bins); 
 \item among non-AGNs, the fraction of HII-like nuclei decreases gradually with increasing density, 
 overall by a factor of four. RET and PAS increase by a factor of two and four respectively. A similar result
 was obtained by Miller et al. (2003).
 \end{itemize}
 The second result need not imply that the local environment has a direct influence on 
 the nuclear activity of galaxies, but partly reflects
 the influence of the  environment on the "extended galaxy" that harbors some nuclear activity, also called
 "morphology segregation" (Dressler, 1980).
 To distinguish the two effects, it is recommended that we normalize the number of
 objects in the various classes of nuclear activity not to the total number of surveyed galaxies 
 (irrespective of the morphological type) but the number of galaxies of certain type classes,
 as in Fig. \ref{fracdens}.
 Guided by the distribution of AGNs with respect to the Hubble classification (Fig. \ref{morf}),
 it is apparent that (among AGNs) SEY and sAGN are more frequent in S0a-Sb, while
 LIN are frequent in E-Sbs.
 Among non-AGNs, HII are frequent in all late-type galaxies (including S0a and BCD),
 RET are found among early-type galaxies (i.e. dE,dS0,E,S0+S0a) and PAS in dE,dS0,E,S0.
 Fig. \ref{fracdens} allows us to study the nuclear activity of galaxies irrespective of "morphology segregation".
 In other words, Fig. \ref{dens} includes the effects of the environment on the galaxy as a whole (morphology)
 and on the nuclear region,  while  Fig. \ref{fracdens} emphasizes the effects of the environment
 on the nuclear region alone.

In conclusion  (see Fig. \ref{fracdens}): 
 \begin{itemize}
 \item The fraction of AGNs associated with both late- and early-type giant galaxies
 is lower at most by a factor of two in the densest galaxy environment.
 \item The fraction of star-forming (HII) galaxies is lower by a factor of at most two in the densest galaxy environment.
 \item The fraction of ex-star-forming (RET) galaxies decreases gradually by a factor of two with increasing density 
 from the lowest to the highest density bin.
 \item The fraction of passive (PAS) galaxies increases gradually by a factor of two with increasing density 
 from the lowest to the highest density bin.
 \item All signs of nuclear activity, either associated with a central black hole or with  star formation,
 are significantly lower in the cores of the rich clusters than lower density environments.
 \end{itemize}
\section{Discussion and conclusion}
 \label{Discussion}
\begin{figure}
 \begin{center}
\includegraphics[width=9cm]{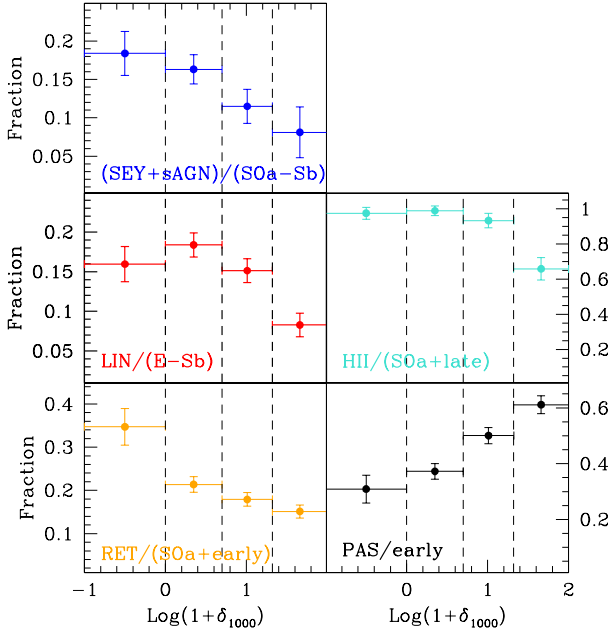}
\caption{The frequency distribution of the ratio of SEY+sAGN to giant spirals (SOa-Sb); 
of LIN to giant galaxies (E-Sb); of RET to early-type galaxies (dE-S0a); of HII to SOa+late-type galaxies; 
and of PAS to early-type (dE-S0) galaxies in four bins of local galaxy density. }
\label{fracdens}
\end{center}
\end{figure}
 A dependence of AGN activity on local galaxy density similar to the one found in this work was proposed by
 Mahajan et al. (2010), who analyzed almost exactly the same sky volume.
 This apparent agreement is however the consequence of a manifold blunder in their analysis.
 First of all, the galaxy sample in the Coma Supercluster used by them suffers from 20\% incompleteness at high luminosity
 (see Section 2) owing to the incompleteness of the SDSS spectral database, which is entirely disregarded in the 
 Mahajan et al. (2010) analysis.\\
 Secondly the  $\rm H\alpha$ spectral line measurements from the SDSS database have not been corrected
 for underlying stellar absorption.  
 Thirdly the fraction of AGNs is computed by Mahajan et al. (2010) by normalizing to the total 
 number of galaxies, irrespective of their morphology,
 Owing to the first omission and the adoption of the BPT diagnostic, Mahajan et al. (2010)
 failed to classify as AGNs an enormous number of objects with [NII] in emission and $\rm H\alpha$ in
 weak absorption, particularly ones present in the core of rich clusters, 
 that would increase the frequency of AGNs in the densest environments.
 At the same time, they overestimated the number of AGNs among galaxies with [NII] and $\rm H\alpha$ in emission
 (that become HII like nuclei when corrected). These are predominantly found in the field.
 If Mahajan et al. (2010) had adopted the correct procedure in producing their BPT diagram, 
 they would have concluded that the fraction of AGNs would
 increase, not decrease, with galaxy density. 
 The reason why AGNs are missing in clusters is that most galaxies that are bona-fide AGNs according to the BPT diagnostic
 diagram become fake-AGNs (RET) according to the WHAN diagnostic diagram, making the whole issue
 of the influence of the environment on the formation of AGNs very slippery.\\
 Using the AGN classification scheme of Kauffmann et al. (2003) based on their revised BPT diagram,
 Kauffmann et al. (2004) analyzed the properties of powerful AGNs among $\sim 122000$ galaxies with $14.5<r<17.77$ 
 and median redshift $z \sim 0.1$, from the SDSS DR1. The subtraction of the underlying stellar continuum 
 was performed by fitting the emission-line-free regions of the spectrum with a model galaxy spectrum.  
 They concluded that powerful AGNs occur predominantly in massive galaxies with 
 $>10^{10} ~M_\odot$, another manifestation of "downsizing" (Cowie et al. 1996, Gavazzi et al. 1996, Fontanot et al. 2009).
 Powerful AGNs are found among late-type, blue, star-forming galaxies.
 The decrease in their frequency with increasing galaxy density is mostly due to 
 the decrease in the frequency of galaxies that are able to host AGNs (morphology segregation),
 but - to a lesser extent - is due to a genuine extra decrease in AGN frequency in high density environments.
 Von der Linden et al. (2010) argued that galaxies in dense environments are less likely to 
 host a powerful optical AGN or a star-forming nuclear region. 
 The frequency of optical AGNs in early-type galaxies declines in high density environments 
 (by approximately a factor of two between the field and the center of clusters). They interpreted 
 this finding by arguing that AGNs are fueled 
 by mass loss from evolved stars. The mean age of the red galaxies increases towards the cluster center and provides 
 less fuel for the central black hole.\\
 Our analysis, which adopts the criteria of Cid Fernandes et al. (2010, 2011) for the classification of nuclei,
 is in full agreement with both Kauffmann et al. (2004) and Von der Linden et al. (2010).
 We conclude that, 
 even after eliminating the effects of morphology segregation
 (by counting the frequency of nuclear activity in the proper bins of Hubble type),
 there is a residual decline in the frequency of nuclear 
 activity (either triggered by black holes or sustained by the star formation) with increasing galaxy density 
 (see Fig. \ref{fracdens}).
 At any luminosity, this decline is quantified in  
 a factor of $\sim$ 2 per interval of density contrast of approximately 20, 
 going from the less dense filaments that form the local cosmic web
 to the densest cores of rich clusters.
 Whichever mechanism is responsible for the suppression of over star formation in galaxies, and the
 migration of galaxy morphologies toward earlier Hubble types in rich and dense environments 
 (see Boselli \& Gavazzi 2006, Blanton \& Moustakas 2009, Fumagalli \& Gavazzi 2008), 
 it has far more severe consequences for their nuclear regions.
 We are not yet in the position to distinguish these two effects.  
 However, we are in the process of obtaining a complete set of $\rm H\alpha$ imaging follow-up observations 
 of the Coma Supercluster that is
 appropriate for distinguishing the nuclear from the extended star-formation in galaxies in this representative part of the local Universe.\\

\section{Acknowledgments}
G. Gavazzi wishes to thank his first year master students who assisted during the observing nights at Loiano for many years.
We also thank R. Gualandi and all night assistants of the Loiano Observatory, and the coordinator of the TAC,
 V. Zitelli for the generous time allocation. We thank O. Cucciati for computing the local galaxy density parameter.
We are grateful to P. Franzetti and A. Donati for their contribution to GoldMine, 
the Galaxy On Line Database extensively used in this work (http://goldmine.mib.infn.it).
We acknowledge the constructive criticism from an anonymous referee.\\
The present study could not have been conceived without the DR7 of SDSS. 
Funding for the Sloan Digital Sky Survey (SDSS) and SDSS-II has been provided by the 
 Alfred P. Sloan Foundation, the Participating Institutions, the National Science Foundation, 
 the U.S. Department of Energy, the National Aeronautics and Space Administration, 
 the Japanese Monbukagakusho, and 
 the Max Planck Society, and the Higher Education Funding Council for England. 
 The SDSS Web site is http://www.sdss.org/.
 The SDSS is managed by the Astrophysical Research Consortium (ARC) for the Participating Institutions. 
 The Participating Institutions are the American Museum of Natural History, Astrophysical Institute Potsdam, 
 University of Basel, University of Cambridge, Case Western Reserve University, The University of Chicago, 
 Drexel University, Fermilab, the Institute for Advanced Study, the Japan Participation Group, 
 The Johns Hopkins University, the Joint Institute for Nuclear Astrophysics, the Kavli Institute for 
 Particle Astrophysics and Cosmology, the Korean Scientist Group, the Chinese Academy of Sciences (LAMOST), 
 Los Alamos National Laboratory, the Max-Planck-Institute for Astronomy (MPIA), the Max-Planck-Institute 
 for Astrophysics (MPA), New Mexico State University, Ohio State University, University of Pittsburgh, 
 University of Portsmouth, Princeton University, the United States Naval Observatory, and the University 
 of Washington.
 G. Gavazzi acknowledges financial support from Italian MIUR PRIN contract 200854ECE5.

\clearpage
\begin{onecolumn}
\scriptsize
\tiny
\begin{longtable}{lcccccccc}
 \caption{Parameters of 283 galaxies observed at Loiano.}\\
\hline
\noalign{\smallskip}
  Gal & RA    &  Dec      &    z  & $EW \rm H\alpha$  &   $EW[NII2]$   &       Class  &   Av.sp.  &   Activity \\
      & J2000 &  J20000   &       &  $\AA$   &    $\AA$  &         &       	  &		\\
  (1) &   (2) &  (3)      & (4)   & (5)      &    (6)    &    (7)  &        (8)   & (9)        \\
\hline
\noalign{\smallskip}
  CGCG522-004        & 01:47:16.2 & +35:33:48 & 0.01559   & -0.554  & -2.798   & LIN  &   -	    &	 -	  \\ 
  CGCG522-093        & 01:58:35.2 & +38:43:07 & 0.01687   & -9.228  & -5.839   &sAGN  &   -	    &	 -	  \\ 
  CGCG522-104        & 02:00:59.7 & +38:47:05 & 0.01888   & -13.23  & -7.936   &sAGN  &   -	    &	 -	  \\ 
  CGCG522-106        & 02:03:44.8 & +38:15:31 & 0.01922   & -3.166  & -2.952   & LIN  &   -	    &	 -	  \\ 
  NGC0891            & 02:22:33.4 & +42:20:57 & 0.00176   & -35.99  & -15.06   & HII  &  NED	    &	 -	  \\ 
  NGC0925            & 02:27:16.9 & +33:34:45 & 0.00184   & -35.87  & -8.241   & HII  &  NED	    &	 -	  \\ 
  UGC01935           & 02:28:14.5 & +31:18:42 & 0.01665   & -144.8  & -10.65   & SY1  &   -	    &	Sey1	  \\ 
  NGC1068            & 02:42:40.7 & -00:00:48 & 0.00379   & -81.52  & -150.9   & SY1  &  NED	    &	Sey2	  \\ 
  UGC02855           & 03:48:20.7 & +70:07:58 & 0.00400   & -23.25  & -11.18   & HII  &   -	    &	 -	  \\ 
  NGC1507            & 04:04:27.2 & -02:11:19 & 0.00287   & -82.32  & -9.657   & HII  &   -	    &	 -	  \\ 
  NGC1637            & 04:41:28.2 & -02:51:29 & 0.00239   & -37.04  & -23.09   &sAGN  &   -	    &	 -	  \\ 
  NGC1961            & 05:42:04.6 & +69:22:42 & 0.01312   & -10.06  & -19.17   &sAGN  &  NED	    &	LINER	  \\ 
  NGC2273            & 06:50:08.7 & +60:50:45 & 0.00613   & -30.16  & -28.54   & SY1  &  NED	    &	Sey2	  \\ 
  NGC2339            & 07:08:20.5 & +18:46:49 & 0.00735   & -37.87  & -28.26   &sAGN  &  NED	    &	 -	  \\ 
  UGC03809           & 07:27:04.1 & +80:10:41 & 0.00735   &  0.910  & -1.431   & RET  &  NED	    &	 -	  \\ 
  NGC2366            & 07:28:54.7 & +69:12:57 & 0.00026   &  -2339  & -20.87   & HII  &   NED	    &	  -	  \\ 
  NGC2403            & 07:36:51.4 & +65:36:09 & 0.00043   &  -0.97  & -1.402   & LIN  &   NED	    &	  -	  \\ 
  NGC2460            & 07:56:52.3 & +60:20:58 & 0.00481   & -7.404  & -4.505   &sAGN  &   -	    &	 -	  \\ 
  NGC2500            & 08:01:53.2 & +50:44:14 & 0.00168   & -1.484  & -0.099   & HII  &  NED	    &	 -	  \\ 
  NGC2537            & 08:13:14.6 & +45:59:23 & 0.00143   & -3.606  & -2.357   & HII  &   -	    &	HII	  \\ 
  UGC04278           & 08:13:58.9 & +45:44:32 & 0.00180   & -129    & -10      & HII  &  SDSS	    &	 -	  \\ 
  CGCG119-016        & 08:14:14.2 & +21:21:20 & 0.01129   & -0.319  & -2.227   & LIN  &   -	    &	 -	  \\ 
  CGCG119-029        & 08:17:25.8 & +21:41:08 & 0.01189   & -9.690  & -5.93    &sAGN  &   -	    &	 -	  \\ 
  CGCG119-035        & 08:17:56.8 & +22:26:08 & 0.00699   & -53.40  & -6.424   & HII  &  SDSS	    &	 -	  \\ 
  NGC2549            & 08:18:58.4 & +57:48:11 & 0.00346   &  1.885  & -0.449   & PAS  &  NED	    &	 -	  \\ 
  CGCG119-055        & 08:19:38.0 & +21:06:51 & 0.01882   & -25.72  & -23.6    &sAGN  & -	    &	 -        \\
  CGCG119-056        & 08:19:41.3 & +22:02:31 & 0.01151   & -48.29  & -9.248   & HII  &  SDSS	    &	 -	  \\ 
  CGCG119-057        & 08:19:48.3 & +22:01:53 & 0.01194   &  1.481  & -1.812   & RET  &   -	    &	 -	  \\ 
  CGCG119-066        & 08:20:49.3 & +22:39:28 & 0.01381   & -21.90  & -8.39    & HII  &  SDSS	    &	 -	  \\ 
  CGCG119-085        & 08:24:20.2 & +20:32:00 & 0.01991   & -9.358  & -9.075   &sAGN  &  SDSS	    &	 -	  \\ 
  CGCG119-096        & 08:26:29.2 & +22:15:40 & 0.02834   & -39.12  & -14.07   & HII  &  SDSS	    &	 -	  \\ 
  CGCG119-109        & 08:27:41.0 & +21:28:47 & 0.01447   & -6.181  & -3.601   & HII  & -	    &	 -        \\
  NGC2648            & 08:42:39.8 & +14:17:08 & 0.00687   & 0.6825  & -2.224   & RET  &   -	    &	 -	  \\ 
  NGC2654            & 08:49:11.9 & +60:13:16 & 0.00449   & 0.0     & -1.235   & RET  &  SDSS	    &	 -	  \\ 
  NGC2683            & 08:52:41.3 & +33:25:18 & 0.00137   & -1.439  & -2.763   & LIN  &  NED	    &	 -	  \\ 
  NGC2685            & 08:55:34.7 & +58:44:04 & 0.00294   & 0.742   & -1.323   & RET  &  NED	    &	Sey2	  \\ 
  UGC4881S           & 09:15:54.0 & +44:19:49 & 0.03948   & -36.35  & -19.07   &sAGN  & -	    &	 -        \\    
  UGC4881N           & 09:15:55.0 & +44:19:58 & 0.03974   & -24.8   & -15.72   &sAGN  & -	    &	 -        \\    
  NGC2841            & 09:22:02.6 & +50:58:35 & 0.00212   & -1.045  & -3.727   & LIN  &  NED	    &	LINER	  \\ 
  NGC2903            & 09:32:10.1 & +21:30:03 & 0.00185   & -52.78  & -19.8    & HII  &  NED	    &	 -	  \\ 
  NGC3031            & 09:55:33.2 & +69:03:55 & -0.0001   & -3.207  & -6.37    & LIN  &  NED	    &	 -	  \\ 
  NGC3115            & 10:05:14.0 & -07:43:07 & 0.00221   &  0.0    &  0.0     & PAS  &  NED	    &	 -	  \\ 
  CGCG123-035        & 10:17:39.7 & +22:48:36 & 0.00391   & -10.12  & -4.748   & HII  &   -	    &	 -	  \\ 
  CGCG124-003        & 10:20:38.0 & +25:30:17 & 0.02002   &  0.923  &  0.0     & PAS  & -	    &	-     \\    
  IC0610             & 10:26:28.4 & +20:13:41 & 0.00390   & -6.495  & -2.79    & HII  &   -	    &	 -	  \\ 
  CGCG124-033        & 10:31:18.7 & +25:51:13 & 0.01950   &  3.353  & 0.0      & PAS  &   -	    &	 -	  \\ 
  NGC3287            & 10:34:47.3 & +21:38:54 & 0.00435   & -16.56  & -7.103   & HII  &  SDSS	    &	 -	  \\ 
  CGCG154-030        & 10:36:16.0 & +26:57:43 & 0.02134   &  1.563  &  0.0     & PAS  & -	    &	-     \\    
  CGCG125-004        & 10:45:32.3 & +24:09:00 & 0.02043   & -7.33   & -4.148   & HII  &   -	    &	 -	  \\ 
  CGCG155-023        & 10:50:22.7 & +26:44:06 & 0.02168   & -11.81  & -3.975   & HII  &   -	    &	 -	  \\ 
  NGC3412            & 10:50:53.3 & +13:24:44 & 0.00280   &  1.67   & 0.0      & PAS  &  NED	    &	 -	  \\ 
  NGC3424            & 10:51:46.3 & +32:54:03 & 0.00498   & -19.56  & -13.84   &sAGN  &  SDSS	    &	 -	  \\ 
  NGC3437            & 10:52:35.7 & +22:56:03 & 0.00428   & -27.35  & -11.03   & HII  &  SDSS	    &	 -	  \\ 
  CGCG155-035        & 10:54:20.9 & +27:14:23 & 0.00445   & -7.90   & -2.866   & HII  &   -	    &	 -	  \\ 
  NGC3457            & 10:54:48.6 & +17:37:16 & 0.00386   & 1.389   & -1.007   & RET  &   -	    &	 -	  \\ 
  105825+241335      & 10:58:25.2 & +24:13:35 & 0.02145   & -1.527  & -3.268   & LIN  &   -	    &	 -	  \\ 
  NGC3485            & 11:00:02.4 & +14:50:30 & 0.00479   & -10.05  & -5.731   &sAGN  &   -	    &	 -	  \\ 
  110127+274310      & 11:01:27.8 & +27:43:10 & 0.02970   & 2.37    & -3.39    & RET  &   -	    &	 -	  \\ 
  CGCG267-037        & 11:03:11.0 & +56:13:18 & 0.00507   & 0.848   & -0.912   & RET  &   -	    &	 -	  \\ 
  NGC3512            & 11:04:02.9 & +28:02:13 & 0.00459   & -1.319  & -2.938   & LIN  &  SDSS	    &	 -	  \\ 
  CGCG155-060        & 11:05:24.0 & +26:21:53 & 0.02199   & 0.906   & -0.745   & RET  & -	    &	-     \\    
  CGCG125-036        & 11:09:54.5 & +24:15:25 & 0.02097   & -56.82  & -17.04   & HII  &   -	    &	 -	  \\ 
  NGC3596            & 11:15:06.2 & +14:47:13 & 0.00397   & -13.7   & -6.386   & HII  &  NED	    &	 -	  \\ 
  NGC3599            & 11:15:27.0 & +18:06:37 & 0.00277   & -0.315  & -1.2     & LIN  &   -	    &	 -	  \\ 
  CGCG096-038        & 11:23:45.5 & +17:49:07 & 0.00428   & -30.55  & -9.221   & HII  &   -	    &	 -	  \\ 
  NGC3657            & 11:23:55.6 & +52:55:16 & 0.00405   & -4.844  & -2.347   & HII  &   -	    &	 -	  \\ 
  NGC3683            & 11:27:31.8 & +56:52:37 & 0.00572   & -9.177  & -3.803   & HII  &  SDSS	    &	 -	  \\ 
  CGCG126-070        & 11:29:51.0 & +24:56:16 & 0.02515   & -2.233  & -1.356   & HII  & -	    &	-     \\    
  113220+223748      & 11:32:20.0 & +22:37:48 & 0.03049   & -25.87  & -10.59   & HII  & -	    &	-     \\    
  113507+201303      & 11:35:07.0 & +20:13:03 & 0.03119   & -15.68  & -10.2    &sAGN  & -	    &	-     \\    
  NGC3758E           & 11:36:29.0 & +21:35:45 & 0.02850   & -186.9  & 12.08    & SY1  & -	    &	-     \\    
  NGC3758W           & 11:36:29.0 & +21:35:45 & 0.02948   & -22.31  & -13.09   &sAGN  & -	    &	-     \\    
 \caption{Continue}\\
\hline
\noalign{\smallskip}
  Gal & RA    &  Dec      &    z  & $EW \rm H\alpha$  &   $EW[NII2]$  &   Class &   Av.sp.  &   Activity \\
      & J2000 &  J20000   &       &  $\AA$   &    $\AA$  &         &            &            \\
  (1) &   (2) &  (3)      & (4)   & (5)      &    (6)    &    (7)  &      (8)   & (9)        \\
\hline
\noalign{\smallskip}
  CGCG097-023        & 11:36:51.0 & +20:00:17 & 0.02110   & 1.611   & 0.0      & PAS & -	   &   -     \\    
  CGCG097-026        & 11:36:54.4 & +19:58:15 & 0.02064   & -46.71  & -18.09   & HII &   -	   &	-	 \\ 
  CGCG127-006        & 11:37:43.0 & +22:00:34 & 0.02982   & 1.574   & -1.428   & RET & -	   &   -     \\    
  CGCG127-007        & 11:37:49.0 & +22:01:33 & 0.03029   & 1.022   & -2.654   & RET & -	   &   -     \\    
  CGCG127-012S       & 11:37:53.7 & +21:58:51 & 0.02911   & -1.472  & -2.095   & LIN &   -	   &	-	 \\ 
  NGC3773            & 11:38:12.9 & +12:06:43 & 0.00327   & -91.57  & -15.93   & HII &  NED	   &   Starb	 \\ 
  FGC1287            & 11:39:11.0 & +19:35:09 & 0.02279   & -13.49  & -6.852   & HII & -	   &   -     \\    
  CGCG097-044        & 11:39:47.5 & +19:56:00 & 0.03646   & -5.509  & -6.402   &sAGN &   -	   &	-	 \\ 
  113959+200935      & 11:39:59.0 & +20:09:35 & 0.02395   & 0.906   & 0.0      & PAS & -	   &   -     \\    
  CGCG157-012        & 11:40:39.0 & +28:51:39 & 0.02287   & -17.85  & -5.929   & HII & -	   &   -     \\    
  CGCG127-025S       & 11:40:44.2 & +22:25:46 & 0.02360   & -29.48  & -13.64   & HII &   -	   &	-	 \\ 
  CGCG127-025N       & 11:40:44.6 & +22:26:49 & 0.02353   & -39.82  & -19.58   & HII &  SDSS	   &	-	 \\ 
  CGCG127-032        & 11:42:09.1 & +20:18:56 & 0.01922   & 1.214   & -1.466   & RET &   -	   &	-	 \\ 
  UGC06678           & 11:43:01.9 & +26:15:30 & 0.03161   & 1.375   & -0.580   & RET &  SDSS	   &	-	 \\ 
  GCG097-082         & 11:43:24.6 & +19:44:59 & 0.02034   & 2.273   & 0.0      & PAS &   -	   &	-	 \\ 
  CGCG097-092        & 11:43:58.2 & +20:11:06 & 0.02125   & -77.24  & -21.96   & HII &   -	   &	-	 \\ 
  CGCG097-091        & 11:43:59.0 & +20:04:37 & 0.02459   & -20.28  & -6.339   & HII &  SDSS	   &	-	 \\ 
  CGCG097-102N       & 11:44:17.2 & +20:13:24 & 0.02124   &  0.0    &  0.0     & PAS &   -	   &	-	 \\ 
  CGCG097-121        & 11:44:47.0 & +20:07:30 & 0.02195   & -0.962  & -2.665   & LIN &   -	   &	-	 \\ 
  CGCG097-114        & 11:44:47.8 & +19:46:24 & 0.02766   & -55.38  & -16.19   & HII &  SDSS	   &	-	 \\ 
  114448+194828      & 11:44:48.9 & +19:48:28 & 0.06770   & -18.78  & -8.65    & HII &  SDSS	   &	-	 \\ 
  CGCG097-120        & 11:44:49.2 & +19:47:42 & 0.01866   & -9.627  & -6.969   &sAGN &   -	   &	-	 \\ 
  CGCG097-122        & 11:44:52.2 & +19:27:15 & 0.01823   & -31.86  & -11.2    & HII &   -	   &	-	 \\ 
  CGCG097-128        & 11:45:03.9 & +19:37:14 & 0.02171   &  0.0    &  0.0     & PAS &   -	   &	-	 \\ 
  CGCG097-127        & 11:45:05.0 & +19:36:23 & 0.02171   & -0.370  & -2.569   & LIN &  SDSS	   &   AGN	 \\ 
  CGCG097-129E       & 11:45:07.0 & +19:58:01 & 0.02517   & -5.977  & -6.482   &sAGN &   -	   &	-	 \\ 
  CGCG097-130        & 11:45:15.0 & -19:23:31 & 0.02230   & 0.0     & 0.0      & PAS & -	   &   -     \\    
  CGCG097-133E       & 11:45:17.6 & +20:01:10 & 0.04845   & -23.34  & -8.531   & HII &   -	   &	-	 \\ 
  CGCG097-134        & 11:45:30.0 & +19:24:00 & 0.02483   & 1.966   & 0.0      & PAS & -	   &   -     \\    
  CGCG097-139        & 11:45:49.5 & +19:46:03 & 0.02320   &  0.0    &  0.0     & PAS &   -	   &	-	 \\ 
  CGCG127-051N       & 11:45:59.4 & +20:26:49 & 0.02370   & -26.34  & -11.14   & HII &   -	   &	-	 \\ 
  CGCG127-051S       & 11:45:59.9 & +20:26:20 & 0.02438   & -35.2   & -17.33   & HII &   -	   &	-	 \\ 
  CGCG127-054        & 11:46:47.2 & +20:40:32 & 0.02343   & 0.0     & -2.258   & RET &  SDSS	   &	-	 \\ 
  114658+194444      & 11:46:58.0 & +19:44:44 & 0.02073   & 1.345   & -1.358   & RET & -	   &   -     \\    
  CGCG157-032        & 11:47:07.9 & +29:34:39 & 0.02271   &  0.0    &  0.0     & PAS &  SDSS	   &	-	 \\ 
  CGCG097-152        & 11:47:39.3 & +19:56:22 & 0.02059   &  0.0    &  0.0     & PAS &  SDSS	   &	-	 \\ 
  CGCG157-035        & 11:48:03.4 & +30:21:34 & 0.02101   & -10.54  & -4.672   & HII &  SDSS	   &	-	 \\ 
  114845+293828      & 11:48:45.0 & +29:38:28 & 0.02237   & -8.484  & -11.03   & SY1 & -	   &   -     \\    
  CGCG127-063        & 11:49:59.3 & +21:20:01 & 0.02624   & 0.5     & -0.73    & RET &   -	   &	-	 \\ 
  CGCG127-073        & 11:51:02.3 & +20:47:59 & 0.02143   & 0.500   & -1.658   & RET &   -	   &	-	 \\ 
  CGCG127-075        & 11:51:21.0 & +21:53:21 & 0.02648   &  0.0    &  0.0     & PAS &   -	   &	-	 \\ 
  CGCG127-076W       & 11:51:26.6 & +22:01:41 & 0.02834   & 1.401   & 0.0      & PAS &   -	   &	-	 \\ 
  CGCG127-076E       & 11:51:28.2 & +22:01:33 & 0.02566   &  0.0    &  0.0     & PAS &   -	   &	-	 \\ 
  CGCG157-045N       & 11:51:32.1 & +27:38:49 & 0.02891   &  0.0    &  0.0     & PAS &  SDSS	   &	-	 \\ 
  CGCG157-045S       & 11:51:34.9 & +27:38:23 & 0.02901   &  0.0    &  0.0     & PAS &   -	   &	-	 \\ 
  CGCG127-086        & 11:52:36.0 & +23:34:55 & 0.02289   & 0.920   & 0.0      & PAS & -	   &   -     \\    
  CGCG127-089        & 11:52:46.0 & +20:59:20 & 0.02087   & 1.309   & 0.0      & PAS & -	   &   -     \\    
  CGCG127-092        & 11:53:10.0 & +20:39:28 & 0.02333   & -1.583  & -6.848   & LIN & -	   &   -     \\    
  CGCG127-094        & 11:53:17.0 & +23:27:52 & 0.02479   & -27.27  & -9.082   & HII &   -	   &	-	 \\ 
  115348+250912      & 11:53:48.1 & +25:09:12 & 0.02837   & -28.09  & -12.18   & HII &   -	   &	-	 \\ 
  CGCG127-101        & 11:53:57.5 & +25:41:06 & 0.01272   & 1.595   & 0.0      & PAS &   -	   &	-	 \\ 
  CGCG127-110        & 11:57:20.9 & +25:11:43 & 0.01501   & -2.286  & -2.163   & LIN &   -	   &	-	 \\ 
  CGCG127-114W       & 11:57:48.2 & +25:16:14 & 0.01591   & -75.52  & -18.87   & HII &  SDSS	   &	-	 \\ 
  CGCG127-114E       & 11:57:50.0 & +25:16:14 & 0.01591   & -36.81  & -9.359   & HII &   -	   &	-	 \\ 
  CGCG157-064        & 11:57:51.0 & +29:02:20 & 0.02156   & -41.94  & -14.21   & HII & -	   &   -     \\    
  CGCG127-116        & 11:57:59.0 & +23:12:03 & 0.02370   & 0.446   & -0.693   & RET & -	   &   -     \\    
  CGCG157-065        & 11:58:05.2 & +27:52:44 & 0.01123   & -166.0  & -31.74   & HII &  NED+SDSS   &	-	 \\ 
  CGCG157-067        & 11:58:20.7 & +28:21:35 & 0.02827   &  0.0    &  0.0     & PAS &   -	   &	-	 \\ 
  CGCG127-122        & 11:58:42.9 & +25:02:25 & 0.01457   & 0.7803  & -1.702   & RET &   -	   &	-	 \\ 
  CGCG127-130        & 11:59:36.0 & +21:14:49 & 0.02635   & 0.136   & -1.189   & RET & -	   &   -     \\    
  NGC4031            & 12:00:31.0 & +31:56:51 & 0.02607   & -21.34  & -9.448   & HII & -	   &   -     \\    
  CGCG127-134        & 12:01:12.0 & +22:37:01 & 0.02761   & -17.94  & -5.692   & HII & -	   &   -     \\    
  CGCG013-046        & 12:02:42.2 & +01:58:37 & 0.00659   & -13.95  & -9.624   &sAGN &  NED+SDSS   &	-	 \\ 
  CGCG098-030        & 12:02:50.2 & +18:00:56 & 0.01605   & -46.66  & -18.55   & HII &   -	   &	-	 \\ 
  CGCG128-003        & 12:03:27.0 & +22:12:36 & 0.02160   & -27.95  & -9.976   & HII & -	   &   -     \\    
  NGC4067            & 12:04:11.5 & +10:51:16 & 0.00805   & -0.269  & -1.121   & LIN &   -	   &	-	 \\ 
  CGCG098-046        & 12:04:32.5 & +20:12:18 & 0.02072   & -0.974  & -4.754   & LIN &  SDSS	   &	-	 \\ 
  CGCG098-058        & 12:07:03.5 & +18:31:54 & 0.02404   & 0.0     & -3.099   & RET &  SDSS	   &	-	 \\ 
  CGCG041-041        & 12:07:37.2 & +02:41:26 & 0.00436   & -82.8   & -20.54   & HII &   -	   &	-	 \\ 
  CGCG041-042        & 12:08:11.1 & +02:52:42 & 0.00442   & -113.2  & -60.62   &sAGN &  NED+SDSS   &   Starb	 \\ 
  CGCG128-038        & 12:08:15.3 & +25:48:41 & 0.02234   &  0.0    &  0.0     & PAS &   -	   &	-	 \\ 
  CGCG098-074        & 12:09:58.0 & +19:55:04 & 0.02580   & -14.88  & -8.004   & HII & -	   &   -     \\    
 \caption{Continue}\\
\hline
\noalign{\smallskip}
  Gal & RA    &  Dec      &    z  & $EW \rm H\alpha$  &   $EW[NII2]$  &   Class &   Av.sp.  &   Activity \\
      & J2000 &  J20000   &       &  $\AA$   &    $\AA$  &         &            &            \\
  (1) &   (2) &  (3)      & (4)   & (5)      &    (6)    &    (7)  &      (8)   & (9)        \\
\hline
\noalign{\smallskip}
  CGCG158-036        & 12:10:18.3 & +26:25:51 & 0.02174   &  0.0    &  0.0     & PAS &  SDSS	   &	-	 \\ 
  CGCG158-039        & 12:11:31.0 & +29:05:20 & 0.02625   & 1.097   & 0.0      & PAS & -	   &   -     \\    
  CGCG128-051        & 12:11:52.5 & +24:07:25 & 0.00856   & -9.625  & -6.584   &sAGN &  NED	   &	-	 \\ 
  VCC0073            & 12:13:03.0 & +07:02:20 & 0.00699   & -2.612  & -6.949   & LIN &   -	   &	-	 \\ 
  VCC0092            & 12:13:48.3 & +14:54:01 & -0.0004   & -3.916  & -5.802   & LIN &  NED	   &   LINER	 \\ 
  VCC0097            & 12:13:53.6 & +13:10:22 & 0.00827   & -14.82  & -8.86    &sAGN &  SDSS	   &	-	 \\ 
  VCC0099            & 12:14:02.2 & +06:43:24 & 0.00815   & -5.344  & -2.127   & HII &  SDSS	   &	-	 \\ 
  VCC0120            & 12:14:38.6 & +05:48:21 & 0.00688   & -29.58  & -9.817   & HII &   -	   &	-	 \\ 
  CGCG158-055        & 12:15:59.2 & +27:26:32 & 0.02551   & -12.9   & -7.803   &sAGN &   -	   &	-	 \\ 
  VCC0199            & 12:16:33.8 & +07:27:44 & 0.00868   &  0.0    &  0.0     & PAS &  SDSS	   &	-	 \\ 
  VCC0362            & 12:19:42.3 & +05:32:18 & 0.00512   & -1.673  & -2.282   & LIN &   -	   &	-	 \\ 
  VCC0382            & 12:19:56.1 & +05:20:36 & 0.00793   & -74.39  & -26.45   & HII &  NED	   &	-	 \\ 
  NGC4287            & 12:20:48.5 & +05:38:24 & 0.00718   & -22.1   & -8.184   & HII &   -	   &	-	 \\ 
  CGCG158-084        & 12:21:01.0 & +27:53:52 & 0.02480   & 0.0     & 0.0      & PAS & -	   &   -     \\    
  VCC0449            & 12:21:02.3 & +03:43:20 & 0.00873   & -3.824  & -1.758   & HII &   -	   &	-	 \\ 
  VCC0492            & 12:21:41.5 & +05:23:05 & 0.00770   & -3.436  & -3.098   & LIN &  SDSS	   &	-	 \\ 
  CGCG128-082        & 12:21:45.6 & +25:53:05 & 0.02304   &  0.0    & -2.268   & RET &  SDSS-corr  &	-	 \\ 
  VCC0562            & 12:22:35.9 & +12:09:29 & 0.00014   &  0.0    &  0.0     & PAS &  SDSS	   &	-	 \\ 
  VCC0567            & 12:22:39.0 & +06:40:37 & 0.00788   & -19.05  & -1.218   & HII &  SDSS	   &	-	 \\ 
  VCC0656            & 12:23:38.7 & +06:57:15 & 0.00338   & -1.838  & -1.939   & LIN &  SDSS	   &	-	 \\ 
  VCC0697            & 12:24:05.5 & +07:02:29 & 0.00410   & -15.52  & -5.061   & HII &  SDSS	   &	-	 \\ 
  VCC0792            & 12:25:22.2 & +10:01:01 & 0.00321   & -1.133  & -1.827   & LIN &  NED+SDSS   &	-	 \\ 
  CGCG158-106        & 12:26:19.9 & +26:50:18 & 0.02361   &  0.0    &  0.0     & PAS &   -	   &	-	 \\ 
  VCC0912            & 12:26:32.2 & +12:36:40 & 0.00034   & -45.58  & -14.27   & HII &  SDSS	   &	-	 \\ 
  VCC0958            & 12:26:56.4 & +15:02:51 & -0.0008   & -3.428  & -5.434   & LIN &  NED	   &	-	 \\ 
  CGCG128-089        & 12:26:59.5 & +22:38:23 & 0.02281   &  0.441  & -1.907   & RET &  SDSS	   &	-	 \\ 
  CGCG158-112        & 12:27:50.3 & +26:59:37 & 0.02390   & -4.337  & -5.01    & LIN &  SDSS	   &	-	 \\ 
  CGCG159-002        & 12:28:15.4 & +28:37:13 & 0.00220   & -8.183  & -4.646   & HII &  NED	   &	-	 \\ 
  VCC1110            & 12:28:29.6 & +17:05:06 & 0.00651   & -0.413  & -3.252   & LIN &  NED	   &   LINER	 \\ 
  NGC4460            & 12:28:45.6 & +44:51:51 & 0.00163   & -34.77  & -7.781   & HII &  NED	   &	-	 \\ 
  VCC1145            & 12:28:59.0 & +03:34:14 & 0.00294   & -0.965  & -3.249   & LIN &  NED	   &	-	 \\ 
  NGC4490            & 12:30:36.2 & +41:38:38 & 0.00188   & -10.55  & -2.92    & HII &  NED	   &	-	 \\ 
  VCC1330            & 12:30:59.7 & +08:04:40 & 0.00580   &  1.616  & -1.427   & RET &   -	   &	-	 \\ 
  CGCG159-010        & 12:31:38.6 & +27:29:45 & 0.02337   & -4.485  & -1.948   & HII &   -	   &	-	 \\ 
  VCC1412            & 12:32:06.2 & +11:10:35 & 0.00447   & 2.314   & 0.0      & PAS &  NED	   &	-	 \\ 
  CGCG014-062        & 12:32:28.2 & +00:23:23 & 0.00503   & -32.35  & 0.0      & HII &  NED+SDSS   &	-	 \\ 
  VCC1442            & 12:32:36.5 & +02:39:41 & 0.00576   & -78.15  & -9.636   & HII &   -	   &	-	 \\ 
  VCC1555            & 12:34:20.3 & +08:11:52 & 0.00655   & -46.91  & -18.96   & HII &  NED	   &	-	 \\ 
  VCC1562            & 12:34:27.0 & +02:11:17 & 0.00603   & -33.47  & -16.19   & HII &  NED	   &	-	 \\ 
  VCC1615            & 12:35:26.5 & +14:29:47 & 0.00162   & 0.364   & -2.348   & RET &  NED	   &   LINER	 \\ 
  CGCG159-024        & 12:35:57.6 & +27:57:36 & 0.00269   & -10.74  & -5.087   & HII &  NED	   &	-	 \\ 
  CGCG129-011        & 12:36:30.3 & +26:11:58 & 0.02377   & 2.246   & -8.812   & RET &   -	   &	-	 \\ 
  VCC1690            & 12:36:49.8 & +13:09:47 & -0.0007   & -5.7    & -8.776   & SY1 &  NED	   &   Sey	 \\ 
  CGCG159-033        & 12:37:21.1 & +28:12:29 & 0.02559   & 0.0     & -4.089   & RET &  SDSS	   &	-	 \\ 
  VCC1727            & 12:37:43.5 & +11:49:05 & 0.00506   & -3.64   & -11.27   & SY1 &  NED	   &   Sey2	 \\ 
  VCC1757            & 12:38:17.9 & +13:06:36 & 0.00593   & -25.83  & -6.683   & HII &  SDSS	   &	-	 \\ 
  CGCG159-043        & 12:39:18.0 & +27:46:22 & 0.02185   & 0.711   & -1.34    & RET & -	   &   -     \\    
  CGCG159-045        & 12:39:18.0 & +28:54:17 & 0.02490   & 1.626   & -2.058   & RET & -	   &   -     \\    
  IC3644             & 12:40:36.2 & +26:30:17 & 0.02212   & -13.98  & -4.167   & HII &   -	   &	-	 \\ 
  IC3646             & 12:40:38.5 & +26:31:34 & 0.02159   & -38.46  & -11.31   & HII &   -	   &	-	 \\ 
  CGCG159-053        & 12:41:08.0 & +29:32:16 & 0.02330   & 0.0     & -1.669   & RET & -	   &   -     \\    
  NGC4615            & 12:41:37.3 & +26:04:22 & 0.01573   & -33.99  & -9.369   & HII &   -	   &	-	 \\ 
  NGC4631            & 12:42:08.0 & +32:32:29 & 0.00202   & -124.2  & -16.25   & HII &  NED+SDSS   &	-	 \\ 
  VCC1923            & 12:42:31.1 & +03:57:37 & 0.00245   & -251.9  & -70.92   & HII &   -	   &	-	 \\ 
  CGCG014-110        & 12:42:32.0 & -00:04:57 & 0.00574   & -36.21  & -13.0    & HII &   -	   &	-	 \\ 
  CGCG129-020        & 12:42:54.0 & +20:59:23 & 0.02182   & -0.316  & -2.462   & LIN &   -	   &	-	 \\ 
  CGCG100-005        & 12:44:53.1 & +18:45:19 & 0.02205   & -29.25  & -16.64   &sAGN &   -	   &	-	 \\ 
  CGCG129-022        & 12:45:11.7 & +23:02:09 & 0.02324   & -0.295  & -1.033   & LIN &   -	   &	-	 \\ 
  CGCG159-070        & 12:45:34.0 & +27:03:35 & 0.02280   & 0.964   & -0.707   & RET & -	   &   -     \\    
  CGCG159-071        & 12:45:43.0 & +29:25:58 & 0.02315   & -34.99  & -11.54   & HII & -	   &   -     \\    
  124708+274736      & 12:47:08.5 & +27:47:36 & 0.02485   & -79.03  & -16.24   & HII &   -	   &	-	 \\ 
  UGC7955            & 12:47:11.0 & +26:42:48 & 0.02236   & -8.625  & -2.832   & HII & -	   &   -     \\    
  CGCG100-007        & 12:47:11.4 & +19:27:52 & 0.02264   & 2.037   & -0.783   & RET &   -	   &	-	 \\ 
  VCC2070            & 12:48:22.9 & +08:29:15 & 0.00336   & -0.47   & -2.432   & LIN &  NED	   &   Sey2	\\ 
  CGCG129-025        & 12:49:34.2 & +25:28:12 & 0.01462   & -7.422  & -2.434   & HII &   -	   &	-	 \\ 
  125209+272838      & 12:52:09.0 & +27:28:38 & 0.01949   & 2.212   & 0.0      & PAS &  SDSS-corr  &	-	 \\ 
  CGCG071-068        & 12:53:50.8 & +09:42:36 & 0.00944   & -56.73  & -23.17   & HII &  SDSS	   &   Starb	 \\ 
  CGCG160-008        & 12:54:19.0 & +27:04:05 & 0.02790   &  0.0    &  0.0     & PAS &   -	   &	-	 \\ 
  CGCG043-068        & 12:55:23.6 & +07:54:34 & 0.00934   & -18.77  & -6.371   & HII &   -	   &	-	 \\ 
  CGCG160-021        & 12:56:12.1 & +27:44:44 & 0.02306   &  0.0    &  0.0     & PAS &   -	   &	-	 \\ 
  CGCG160-025        & 12:56:27.9 & +26:59:14 & 0.02155   & -0.397  & -3.658   & LIN &  SDSS	   &	-	 \\ 
  CGCG160-055        & 12:58:05.6 & +28:14:34 & 0.02351   & -86.04  & -30.69   & HII &  SDSS	   &	-	 \\ 
 \caption{Continue}\\
\hline
\noalign{\smallskip}
  Gal & RA    &  Dec      &    z  & $EW \rm H\alpha$  &   $EW[NII2]$  &   Class &   Av.sp.  &   Activity \\
      & J2000 &  J20000   &       &  $\AA$   &    $\AA$  &         &            &            \\
  (1) &   (2) &  (3)      & (4)   & (5)      &    (6)    &    (7)  &      (8)   & (9)        \\
\hline
\noalign{\smallskip}
  CGCG160-221        & 12:59:13.0 & +27:58:37 & 0.02313   & 1.406   & 0.0      & PAS & -	   &   -     \\    
  CGCG160-225        & 12:59:23.0 & +27:54:39 & 0.02198   & 1.674   & 0.0      & PAS & -	   &   -     \\    
  CGCG130-003        & 12:59:47.3 & +21:48:48 & 0.02387   & -23.69  & -13.64   &sAGN &  SDSS	   &	-	 \\ 
  CGCG160-241        & 13:00:08.1 & +27:58:37 & 0.02166   & 1.099   & 0.0      & PAS &  SDSS	   &	-	 \\ 
  CGCG160-248        & 13:00:17.7 & +27:57:19 & 0.02302   & 0.0     & -1.286   & RET &   -	   &	-	 \\ 
  CGCG043-093        & 13:00:39.1 & +02:30:05 & 0.00320   & -62.39  & -20.17   & HII &  NED+SDSS   &	-	 \\ 
  CGCG160-257        & 13:00:48.8 & +28:09:30 & 0.01961   & -0.434  & -4.896   & LIN &  SDSS	   &	-	 \\ 
  CGCG160-258        & 13:00:51.0 & +28:02:34 & 0.02867   & 1.495   & 0.0      & PAS & -	   &   -     \\    
  CGCG015-055        & 13:00:58.7 & -00:01:39 & 0.00396   & -64.74  & -20.66   & HII &  SDSS	   &   Starb	 \\ 
  CGCG160-261        & 13:00:59.3 & +27:53:59 & 0.02300   & 2.189   & 0.0      & PAS &   -	   &	-	 \\ 
  CGCG160-096S       & 13:01:24.5 & +29:18:30 & 0.02386   & 1.254   & 0.0      & PAS &   -	   &     -	 \\ 
  CGCG160-096N       & 13:01:25.3 & +29:18:50 & 0.02316   & -110.3  & -77.82   & SY1 &  SDSS	   &   Sey2	 \\ 
  CGCG160-095        & 13:01:26.1 & +27:53:09 & 0.01828   & -0.168  & -3.555   & RET &  SDSS	   &	-	 \\ 
  CGCG160-102        & 13:01:43.4 & +29:02:41 & 0.02366   & -7.706  & -5.815   &sAGN &  SDSS	   &	-	 \\ 
  CGCG160-121        & 13:03:29.1 & +26:33:02 & 0.02226   & -3.732  & -2.039   & HII &  SDSS	   &	-	 \\ 
  CGCG130-005        & 13:04:10.8 & +22:17:23 & 0.02386   & -72.05  & -39.34   &sAGN &  SDSS	   &	-	 \\ 
  130553+273307      & 13:05:53.1 & +27:33:08 & 0.01888   & 1.296   & 0.0      & PAS &   -	   &	-	 \\ 
  CGCG160-154        & 13:10:20.0 & +31:26:38 & 0.02540   & 3.687   & -4.801   & RET & -	   &   -     \\    
  CGCG160-163        & 13:12:59.4 & +27:08:30 & 0.05972   & -18.42  & -13.12   &sAGN &   -	   &	-	 \\ 
  CGCG160-166        & 13:13:26.9 & +27:48:09 & 0.02137   & -0.428  & -2.429   & LIN &  SDSS	   &	-	 \\ 
  CGCG130-021        & 13:13:45.3 & +24:58:57 & 0.02418   & -34.63  & -15.94   & HII &   -	   &	-	 \\ 
  CGCG160-173        & 13:16:12.3 & +30:57:01 & 0.01865   & -8.16   & -6.093   &sAGN &   -	   &	-	 \\ 
  CGCG160-177        & 13:16:51.0 & +31:34:52 & 0.02980   & -10.94  & -5.381   & HII & -	   &   -     \\    
  CGCG130-026        & 13:17:19.8 & +20:38:17 & 0.02291   & -10.64  & -7.098   &sAGN &  SDSS	   &	-	 \\ 
  CGCG130-028        & 13:17:28.5 & +20:36:44 & 0.02222   & 0.402   & -2.198   & RET &   -	   &	-	 \\ 
  CGCG160-181        & 13:17:30.6 & +31:05:34 & 0.01851   & -15.01  & -5.528   & HII &   -	   &	-	 \\ 
  CGCG160-182        & 13:17:45.2 & +27:34:12 & 0.02332   & -7.286  & -4.848   &sAGN &  SDSS	   &	-	 \\ 
  CGCG160-198        & 13:20:01.0 & +31:58:53 & 0.01813   &  0.0    &  0.0     & PAS &   -	   &	-	 \\ 
  CGCG160-200        & 13:20:14.0 & +30:53:59 & 0.02448   & -34.14  & -15.18   & HII & -	   &   -     \\    
  CGCG160-202        & 13:20:21.0 & +31:30:53 & 0.01669   & -10.55  & -4.993   & HII & -	   &   -     \\    
  NGC5103            & 13:20:30.1 & +43:05:02 & 0.00425   & 1.232   & -0.522   & RET &   -	   &	-	 \\ 
  CGCG131-003        & 13:21:52.0 & +22:25:45 & 0.03099   & 0.0     & -1.675   & RET &   -	   &	-	 \\ 
  CGCG131-004        & 13:22:26.3 & +21:25:33 & 0.02284   & -42.45  & -25.62   & SY1 &   -	   &	-	 \\ 
  CGCG161-048        & 13:25:57.2 & +31:37:06 & 0.02437   &  0.0    &  0.0     & PAS &  SDSS	   &	-	 \\ 
  CGCG016-069        & 13:26:19.7 & +02:06:03 & 0.00362   & -4.201  & -2.502   & HII &  NED	   &	-	 \\ 
  CGCG161-073        & 13:30:26.0 & +31:37:15 & 0.02441   & -0.306  & -2.201   & LIN &  SDSS	   &	-	 \\ 
  135031+245834      & 13:50:31.0 & +24:58:34 & 0.03041   & 1.048   & -1.834   & RET & -	   &   -     \\    
  CGCG132-023        & 13:51:05.4 & +25:05:37 & 0.02923   &  0.0    &  0.0     & PAS &   -	   &	-	 \\ 
  135244+250223      & 13:52:44.0 & +25:02:23 & 0.02857   & 1.11    & 0.0      & PAS & -	   &   -     \\    
  NGC5372            & 13:54:46.0 & +58:40:01 & 0.00572   & -66.55  & -22.63   & HII &   -	   &	-	 \\ 
  CGCG132-047W       & 13:55:28.4 & +25:04:25 & 0.02955   &  0.0    &  0.0     & PAS &  SDSS	   &	-	 \\ 
  CGCG132-047C       & 13:55:29.8 & +25:04:25 & 0.02954   &  0.0    &  0.0     & PAS &   -	   &	-	 \\ 
  CGCG132-047E       & 13:55:32.6 & +25:04:28 & 0.02850   &  0.0    &  0.0     & PAS &   -	   &	-	 \\ 
  CGCG132-058        & 13:57:15.4 & +24:15:27 & 0.02965   & -2.18   & -1.045   & HII &   -	   &	-	 \\ 
  IC4397             & 14:17:58.7 & +26:24:45 & 0.01474   & -23.3   & -8.77    & HII &   -	   &   HII	 \\ 
  141847+245625      & 14:18:47.0 & +24:56:25 & 0.01766   & -19.35  & -9.173   & HII & -	   &   -     \\    
  CGCG047-022        & 14:21:13.1 & +03:26:09 & 0.00496   & -6.435  & -3.329   & HII &   -	   &	-	 \\ 
  CGCG133-055        & 14:26:17.7 & +26:14:46 & 0.03024   & -5.179  & -4.07    &sAGN &   -	   &	-	 \\ 
  CGCG163-050        & 14:26:59.6 & +26:51:50 & 0.03015   & -10.65  & -3.947   & HII &   -	   &	-	 \\ 
  142816+255055      & 14:28:16.0 & +25:50:55 & 0.01457   & 0.770   & 0.0      & RET & -	   &   -     \\    
  CGCG163-058        & 14:28:31.8 & +27:24:32 & 0.01439   & -3.236  & -5.218   & SY1 &   -	   &	-	 \\ 
  143409+275649      & 14:34:09.0 & +27:56:49 & 0.03174   & 1.743   & -1.857   & RET & -	   &   -     \\    
  CGCG133-088        & 14:34:12.7 & +25:28:05 & 0.01614   & -11.72  & -6.65    &sAGN &   -	   &	-	 \\ 
  CGCG163-085        & 14:35:01.9 & +26:32:38 & 0.03070   & -33.73  & -12.09   & HII &   -	   &	-	 \\ 
  CGCG134-016        & 14:42:18.0 & +25:30:34 & 0.02662   & -3.055  & -3.797   & LIN &   -	   &	-	 \\ 
  UGC9489S           & 14:43:01.0 & +25:08:48 & 0.03083   & -69.68  & -10.7    & HII & -	   &   -     \\    
  UGC9489N           & 14:43:02.0 & +25:09:08 & 0.03088   & -41.77  & -33.8    &sAGN & -	   &   -     \\    
  CGCG164-031        & 14:49:27.0 & +27:46:51 & 0.03097   & -0.877  & -2.039   & LIN & -	   &   -     \\    
  144954+274203      & 14:49:54.3 & +27:42:03 & 0.02973   & -30.69  & -9.636   & HII &   -	   &	-	 \\ 
  CGCG134-054        & 14:54:44.8 & +24:05:49 & 0.01732   & -26.17  & -9.219   & HII &   -	   &	-	 \\ 
  CGCG134-066        & 15:01:53.7 & +25:57:52 & 0.02205   & -2.708  & -5.947   & LIN &   -	   &	-	 \\ 
  IC4530             & 15:03:45.0 & +26:06:03 & 0.03121   & -2.438  & -2.079   & LIN &   -	   &	-	 \\ 
  CGCG165-021        & 15:08:01.0 & +27:58:20 & 0.03196   & 0.767   & -1.769   & RET & -	   &   -     \\    
  NGC5957            & 15:35:23.2 & +12:02:51 & 0.00609   & -1.279  & -3.044   & LIN &  SDSS	   &	-	 \\ 
  CGCG136-042        & 15:39:27.4 & +24:56:51 & 0.02291   & -88.94  & -52.51   &sAGN &   -	   &   Starb	 \\ 
  CGCG136-067        & 15:45:16.0 & +24:37:50 & 0.02233   & 1.482   &  0.0     & PAS & -	   &   -     \\    
  NGC6166            & 16:28:38.5 & +39:33:06 & 0.03035   & -1.369  & -7.51    & LIN &   -	   &	-	 \\ 
   \noalign{\smallskip}
  \hline
  \label{Table4}
  \end{longtable}
  \normalsize
  \end{onecolumn}
  \clearpage
\end{document}